\documentclass[10pt,twocolumn,journal]{IEEEtran}
\usepackage{amssymb,amsfonts}
\usepackage[cmex10]{amsmath}
\usepackage{cite}
\usepackage{pifont}

\usepackage{graphicx}
\usepackage{epstopdf}

\usepackage[mathscr]{eucal}

\newtheorem{lemma}{Lemma}
\newtheorem{theorem}{Theorem}

\newtheorem{corollary}{Corollary}
\newtheorem{defn}{Definition}
\newtheorem{remark}{Remark}
\newtheorem{assum}{Assumption}

\usepackage{enumerate,subfigure}

\def\ve{\varepsilon}
\def\re{\mathop{Re}\nolimits}
\def\diag{\mathop{diag}\nolimits}
\def\R{\mathbb R}
\def\c{\mathbb C}

\def\be{\begin{equation}}
\def\ee{\end{equation}}
\def\ben{\begin{equation*}}
\def\een{\end{equation*}}

\def\sgn{\mathop{sgn}\nolimits}
\def\abs{\mathop{abs}\nolimits}

\newcommand{\dfb}{\stackrel{\Delta}{=}}

\def\E{\mathcal E}
\def\G{\mathcal G}
\def\a{\mathfrak a}

\begin{document}
\title{Opinion Dynamics in Social Networks with Hostile Camps: Consensus vs. Polarization}

\author{Anton V. Proskurnikov, Alexey Matveev and Ming Cao%
\thanks{The work was supported in part by the European Research Council (ERCStG-
307207), St. Petersburg State University, grant 6.38.230.2015 and RFBR, grants 13-08-01014 and 14-08-01015.
Lemma~\ref{lem.main} was obtained by A.S. Matveev under sole support of Russian Science Fund (RSF) grant 14-21-00041 at St. Petersburg State University.
Theorems~\ref{thm.nonlin2} and \ref{thm.nonlin1} were obtained by A.V. Proskurnikov under sole support of RSF 
grant 14-29-00142 at IPME RAS. The results were partly reported on IEEE Multiconference on Systems and Control MSC 2014 (Antibes, France)
and IEEE Conference on Decision and Control CDC 2014 (Los Angeles, CA, USA).}%
\thanks{A.V. Proskurnikov is with ENTEG, Faculty of Mathematics and Natural Sciences, University of Groningen, Groningen,
the Netherlands and also with St. Petersburg State University, ITMO University and Institute
for Problems of Mechanical Engineering of Russian Academy of Sciences (IPME RAS), St.Petersburg, Russia;\;{\tt\small avp1982@gmail.com}}%
\thanks{A. Matveev is with St. Petersburg State University, St. Petersburg, Russia;\;{\tt\small almat1712@yahoo.com}}%
\thanks{M. Cao is with ENTEG, Faculty of Mathematics and Natural Sciences, University of Groningen, Groningen, The Netherlands;\;{\tt\small m.cao@rug.nl}}%
} \maketitle

\begin{abstract}
Most of the distributed protocols for multi-agent consensus assume
that the agents are mutually cooperative and ``trustful'', and so
the couplings among the agents bring the values of their states
closer. Opinion dynamics in social groups, however, require beyond
these conventional models due to ubiquitous competition and distrust
between some pairs of agents, which are usually characterized by
repulsive couplings and may lead to clustering of the opinions. A
simple yet insightful model of opinion dynamics with both attractive
and repulsive couplings was proposed recently by C. Altafini, who
examined first-order consensus algorithms over static signed graphs.
This protocol establishes modulus consensus, where the opinions
become the same in modulus but may differ in signs. In this paper,
we extend the modulus consensus model to the case where the network
topology is an arbitrary time-varying signed graph and prove
reaching  modulus consensus under mild sufficient conditions of
uniform  connectivity of the graph. For cut-balanced graphs, not
only sufficient, but also necessary conditions for modulus consensus
are given.
\end{abstract}


\section{Introduction}

For multi-agent networks, the striking phenomenon of global
consensus  caused by only local interactions  has attracted
long-standing interest from the research community. The interest is
motivated by numerous natural phenomena and engineering designs
related to reaching synchrony or agreement among the agents.
Examples include, but not limited to, intelligence of large
biological populations and multi-robot teams. We refer the reader to
\cite{RenBeardBook,MesbahiEgerBook,RenCaoBook} for excellent surveys
of recent research on consensus protocols and their applications, as
well as historical milestones.

Starting from the DeGroot algorithm of ``iterative pooling''
\cite{DeGroot} for distributed decision making, many consensus
algorithms were based on the principle of \emph{contraction}: every
agent's state constantly evolves to the relative interior of the
convex hull spanned by its own and neighbors' states. Hence, the
convex hull spanned by the states of the agents, driven by such a
protocol, is shrinking over time. Based on the Lyapunov-like
properties of this convex hull \cite{Moro:05,LinFrancis:07} or
relevant results on convergence of infinite products of stochastic
matrices \cite{CaoMorse:08,RenBeardBook,MesbahiEgerBook}, stability
properties of contracting iterations were examined intensively with
special attention on the effect of time-varying interaction
topologies. Necessary and sufficient conditions for consensus under
bidirectional \cite{Blondel:05,Moro:05,MatvPro:2013} and cut-balanced
graphs \cite{TsiTsi:13} boil down to repeated joint connectivity of
the network. For general directed graphs the sufficient condition of
\emph{uniform quasi-strong connectivity} (UQSC) \cite{LinFrancis:07},  is considered to be ``the weakest assumption
on the graph connectivity such that consensus is guaranteed for
arbitrary initial conditions'' \cite{Muenz:11}. This common belief
has recently been confirmed by  results in
\cite{ShiJohansson:13,ShiJohansson:13-1} stating that the UQSC is
necessary and sufficient for robust consensus and consensus with
exponential convergence. Many high-order consensus algorithms either
extend their first-order counterparts \cite{RenBeardBook,
RenCaoBook} or are squarely based on them \cite{Sepul:09}.

Unlike teams of agents that achieve a common goal due to
cooperation, networks where agents can both cooperate and compete
(sometimes referred to as \emph{coopetitive} networks \cite{HuZheng:2014,HuZhu:15}) still demand more thorough mathematically
rigorous analysis. In social networks, competition, antagonism and
distrust between social actors and their groups are ubiquitous
\cite{EasleyKleinberg,WassermanFaust}, which are usually modeled by
\emph{repulsive couplings} or \emph{negative ties}
\cite{Flache:2011} among the agents. A specific example of such
couplings observed in dyadic interactions, is \emph{reactance}
\cite{Dillard:2005} which leads to \emph{boomerang effects}, first
described in \cite{HovlandBook}: in the process of persuasion,
opinions (even close to each other initially) can become opposite.
Analogous phenomenon, referred to as the \emph{group polarization}
\cite{Aronson:2010}, has long been studied in social psychology: the
community divides into two groups, each reaching consensus; the
consensus opinions are not only opposite, but often further away
from each other than the two initial average opinions of the
corresponding groups. Analysis of real-world social networks (e.g.
users of social web-sites \cite{EasleyKleinberg}) shows the strong
correlation between polarization and \emph{structural balance}
\cite{EasleyKleinberg,WassermanFaust} of positive and negative ties.
The latter property implies that community splits into two hostile
camps (e.g. votaries of two political parties), where the relations
inside each faction are cooperative.

It is known that agents' \emph{repulsion} can lead to the clustering
behavior in a complex network \cite{XiaCao:11}. The possibility of
clustering in social groups due to negative ties was demonstrated in
\cite{Flache:2011} (see also references therein); these effects are
still waiting for mathematically rigorous analysis. Most of the
existing works on opinion dynamics focus on the  persistent
disagreement and clustering of opinions caused by bounded confidence
\cite{Krause:2002,DeffuantWeisbuch:2000} or, more generally,
\emph{biased assimilation} \cite{Dandekar:2013}: agents readily
adopt opinions of like-minded neighbors, accepting the ``deviating''
opinions with discretion. In \cite{Altafini:2012,Altafini:2013}
Altafini proposed a simple yet instructive mathematical model of
opinion \emph{polarization} over structurally balanced graphs,
extending conventional first-order consensus algorithms to the case
with antagonistic interactions. These protocols were examined under
the assumption that the interaction graph is static and strongly
connected and shown to establish \emph{modulus consensus}
\cite{MengCaoJohansson:2014}, where the opinions agree in modulus
but may differ in signs. If the graph is structurally balanced, the
modulus of the final opinion is generally non-zero and opinions
either reach consensus or polarize (``bipartite consensus''
\cite{Altafini:2013} is established); otherwise, opinions converge
to zero.

Mathematical examination of polarization behavior due to antagonism
is among the first important steps towards understanding the
dynamics of networks consisting of both cooperative and competitive
agents. Such networks are not confined to social systems; repulsive
interactions play an important role in e.g. motion control of swarms
and other multi-agent formations, where agents may avoid collision
\cite{Romanczuk:12,YuChenCaoLuZhang:13} and distribute evenly on
 circular or other closed curves \cite{WangXieCao:13,Asya:15,Ovchinnikov:14}.
So a number of papers have been published recently studying this
class of networks \cite{Valcher:14,HuZheng:2014,ZhangChen:14,Hendrickx:14,AltafiniLini:15,XiaCaoJohansson:15}.

The aforementioned papers, however, mainly focus on the case where
the interaction topology is static. In the present paper, we
consider Altafini's model on opinion dynamics over general directed
time-varying graphs. Removing the restrictions of static topologies
not only allows one to analyze dynamics of real social networks,
where the agents may change their relationships from friendship to
hostility and vice versa, but also enables one to extend the result
to \emph{non-linear} protocols. In fact we will examine nonlinear
algorithms  in the common framework as linear ones, getting rid of
the restrictions such as monotonicity
\cite{Altafini:2012,Altafini:2013,Smith:1988}.

Our main result states that modulus consensus is  established if the
topology is uniformly \emph{strongly} connected. Unlike cooperative
networks, the uniform strong connectivity cannot be relaxed to the
uniform \emph{quasi-strong} connectivity, which is a commonly
adopted condition for consensus over directed time-varying graph
\cite{LinFrancis:05,Muenz:11}. At the same time, the condition of
uniform strong connectivity is in general not necessary, and filling
the gap between necessary and sufficient conditions remains a tough
problem even in the cooperative case. However, we fill this gap in
the special case of \emph{cut-balanced} graphs, extending necessary
and sufficient consensus criterion from \cite{TsiTsi:13} to modulus
consensus over signed graphs. It should be noticed that results from
\cite{TsiTsi:13} are not directly applicable to signed graphs; in
the special case of bidirectional or ``reciprocal interactions''
they were extended  to the signed case by the \emph{lifting}
technique \cite{Hendrickx:14}. We will make further remarks on this
in the corresponding sections. The results were partly reported
in our conference papers \cite{ProCao:2014,ProMatvCao:2014}.

The paper is organized as follows. Section~\ref{sec.prelim} introduces some preliminary concepts and notations. Section~\ref{sec.setup} gives the setup of the problem in question. Section~\ref{sec.main} presents the main results. Section~\ref{sec.proof} offers the proof of the main results.


\section{Preliminaries}\label{sec.prelim}

Throughout the paper $m:n$, where $m,n$ are integers and $m\le n$,
stands for the  sequence $\{m,m+1,\ldots,n\}$. The sign of a number
$x\in\R$ is denoted by $\sgn x\in\{-1,0,1\}$. The abbreviation
``a.a.'' stands for ``almost all'' (except for the set of zero
Lebesgue measure). Given a matrix $L=(l_{jk})$, let $\abs
L\dfb(|l_{jk}|)$. We also introduce the matrix norm
$|L|_{\infty}\dfb \max_j\sum_k|L_{jk}|$. As usual, for a column
vector $x\in\R^N$, one has $|x|_{\infty}=\max_j |x_j|$ and it is
easily shown that
$|L|_{\infty}=\sup\frac{|Lx|_{\infty}}{|x|_{\infty}}$, where the
supremum is over all column vectors $x\ne 0$ of appropriate
dimensions. Let $\bar 1_N\dfb (1,1,\ldots,1)^T\in\R^N$. Given
$x\in\R$, let $x^+=\max(x,0)$ and $x^-=(-x)^+$, hence $x=x^+-x^-$
and $|x|=x^++x^-$.

\subsection{Signed graphs and their properties}

A (weighted directed) signed graph is a triple $G=(V,E,A)$, where
$V=\{v_1,\ldots,v_N\}$ stands for the set of \emph{nodes}, $E\subset
V\times V$ is a set of \emph{arcs} and $A=(a_{jk})\in\R^{N\times N}$
is a signed \emph{adjacency matrix}, i.e. $a_{jk}\ne 0$ if and only
if $(v_k,v_j)\in E$. Throughout the paper, we confine ourselves to
graphs that have no self-loops ($a_{jj}=0\,\forall j$) and are
\emph{digon sign-symmetric} \cite{Altafini:2013}, i.e. any pair of
opposite arcs (if exists) is identically signed: $a_{jk}a_{kj}\ge
0\,\forall j,k$.
Identifying the set of nodes $V$ with $1:N$, there is a one-to-one
correspondence between signed graphs and their adjacency matrices
$A\in\R^{N\times N}\mapsto G[A]=(1:N,E[A],A)$, where
$E[A]=\{(j,k):a_{kj}\ne 0\}$.

Given $\ve>0$, let $A^{\ve}=(a_{jk}^{\ve})$ stand for the
``truncated'' adjacency matrix: $a^{\ve}_{ij}=a_{ij}$ when
$|a_{ij}|\geq\ve$ and $a^{\ve}_{ij}=0$ otherwise. The corresponding
graph $G^{\ve}\dfb G[A^{\ve}]$ is obtained from $G=G[A]$ by removing
arcs of absolute weight less than $\ve$ and we call it
\emph{$\ve$-skeleton} of the graph $G$.

A \emph{path} connecting nodes $v$ and $v'$ is a sequence of nodes $v_{i_0}:=v,v_{i_1},\ldots,v_{i_{n-1}},v_{i_{n}}:=v'$ ($n\ge 1$) such that $(v_{i_{k-1}},v_{i_{k}})\in E$ for $k\in 1:n$. A path where $v_{i_0}=v_{i_n}$ is referred to as a \emph{cycle}. The cycle is \emph{positive} if $a_{i_0i_1}a_{i_1i_2}\ldots a_{i_{n-1}i_n}>0$ and \emph{negative} otherwise. The digon-symmetric strongly connected graph is structurally balanced if and only if all its oriented cycles are positive \cite{Altafini:2013,EasleyKleinberg}. A node is called \emph{root} if it can be connected with a route to any other node of the graph. A graph is \emph{strongly connected} (SC) if a path between any two different nodes exists.
The graph is \emph{quasi-strongly connected} (QSC) if it has at least one root. Any SC graph is also QSC, each node being a root.
A graph whose $\ve$-skeleton is SC (respectively, QSC) is called \emph{strongly $\ve$-connected} (respectively, \emph{quasi-strongly $\ve$-connected}).

Given a graph $G=(V,E,A)$, its \emph{subgraph} is a graph $G'=(V',E',A')$, where $V'\subseteq V$, $E'\subseteq (V'\times V')\cap E$ and $A'=(a_{ij})_{i,j\in V'}$ stands for the corresponding submatrix of $A$. We call a subgraph \emph{in-isolated} if no arc comes from $V\setminus V'$ to $V'$, i.e. $a_{ji}=0\,\forall i\in V',j\not\in V'$.

We call two disjoint non-empty sets $V_1,V_2\subseteq V$ \emph{hostile camps} in the graph $G$ if $a_{jk}\ge 0$ when $j,k\in V_1$ or $j,k\in V_2$ and $a_{jk}\le 0$ whenever $j\in V_1,k\in V_2$ or
$j\in V_2,k\in V_1$. The graph is \emph{structurally balanced} (SB) \cite{Altafini:2013,EasleyKleinberg} if the set of its nodes can be divided into two hostile camps $V=V_1\cup V_2$.
The digon-symmetric SC graph is structurally balanced if and only if any cycle in it is positive \cite{Altafini:2013,EasleyKleinberg}.

Following \cite{Altafini:2013}, we define the \emph{Laplacian} matrix $L=L[A]$ of the signed graph $G[A]$ as follows
\be\label{eq.laplacian}
L[A]\dfb(L_{jk})_{j,k=1}^N,\quad L_{jk}:=\begin{cases}
-a_{jk},\,j\ne k\\
\sum_{m=1}^N|a_{jm}|, j=k.
\end{cases}
\ee

Equation \eqref{eq.laplacian} is a straightforward extension of the
conventional definition of the Laplacian matrix of a weighted graph
\cite{Murray:04} to the case of signed weights. As implied by the
Gershgorin disk theorem \cite{Altafini:2013}, $L[A]$ has no
eigenvalues in the closed left half-plane
$\bar\c_-=\{\lambda\in\c:\re\lambda\le 0\}$ except for possibly $\lambda=0$.
Unlike the unsigned case, in general $L[A]$ may have no zero
eigenvalue and hence $-L[A]$ may be a Hurwitz matrix. For a SC graph
$G=G[A]$ this is the case if and only if $G$ is not structurally
balanced \cite[Lemma 2]{Altafini:2013}.

\subsection{Some important types of time-varying signed graphs}

Throughout the paper, the term \emph{time-varying (signed) graph}
means the graph $G[A(t)]$, where a time-dependent matrix
$A(t)\in\R^{N\times N}$ is Lebesgue measurable and \emph{locally
bounded}. Given such a graph $G(t)=G[A(t)]$, we say a node $j$ is
\emph{essentially connected} to a node $k$ if
$\int_{t_0}^{\infty}|a_{jk}(s)|ds=\infty$ for some $t_0\ge 0$ (the
latter inequality then holds for any $t_0\ge 0$ since $a_{jk}$ is
locally bounded). Let $\E=\E[A(\cdot)]$ stand for the set of all
such pairs $(j,k)$. Following \cite{MatvPro:2013}, we call the graph
$\G[A(\cdot)]=(1:N,\E)$ \emph{the graph of essential interactions}
and say that the graph $G(\cdot)$ is \emph{essentially strongly
connected} (ESC) if $\G[A(\cdot)]$ is strongly connected. Likewise,
$G(\cdot)$ is \emph{essentially quasi-strongly connected} (EQSC) if
$\G[A(\cdot)]$ is QSC.

The graph $G[A(\cdot)]$ is  said to be \emph{uniformly strongly
connected} (USC) if there exist constants $T>0$ and $\ve>0$ such
that the graph $G[\int_t^{t+T}\abs A(s)\,ds]$ is strongly
$\ve$-connected for any $t\ge 0$. By replacing the word ``strongly''
in the latter definition with ``quasi-strongly'', one defines
\emph{uniformly quasi-strongly connected} (UQSC) time-varying graph.
It may be easily shown that the USC (respectively UQSC) graph is
always ESC (respectively, EQSC), while the inverse is not valid.

The graph $G[A(\cdot)]$  is \emph{cut-balanced}  \cite{TsiTsi:13} if
a constant $K\ge 1$ exists such that for any partition of the nodes
$V'\cup V''=1:N$, $V'\cap V''=\emptyset$, the following inequalities
hold \be\label{eq.cut} K^{-1}\sum_{j\in V'}\sum_{k\in
V''}|a_{kj}|\le \sum_{j\in V'}\sum_{k\in V''}|a_{jk}|\le K\sum_{j\in
V'}\sum_{k\in V''}|a_{kj}|. \ee A typical example of a cut-balanced
graph is the \emph{type-symmetric} graph \cite{TsiTsi:13}, which
means the existence of $K\ge 1$ such that 
\be\label{eq.unibidir}
K^{-1}|a_{kj}(t)|\le |a_{jk}(t)|\le K|a_{kj}(t)|\,\forall t\ge
0\forall j\ne k. \ee
 Other examples include weight-balanced graphs, see \cite{TsiTsi:13} for details. As implied
by \cite[Lemma 1]{TsiTsi:13}, for cut-balanced graphs, the EQSC
property implies ESC; precisely, any quasi-strongly connected
component of the digraph $\G\dfb\G[A(\cdot)]$ is strongly connected,
and a path between $j$ and $k$ exists if and only if the path from
$k$ to $j$ exists.

\section{Problem Setup}\label{sec.setup}

Consider a group of $N\ge 2$ agents indexed $1$ through $N$, the
opinion of the $i$th agent is denoted by $x_i\in\R$ and we define
$x:=(x_1,\ldots,x_N)^T\in\R^N$. The agents update their opinions in
accordance with a distributed protocol as follows:
\be\label{eq.proto0} \dot x(t)=-L[A(t)]x(t),t\ge 0, \ee which can be
written componentwise as \be\label{eq.proto} \dot
x_j(t)=\sum_{k=1}^N|a_{jk}(t)|(x_k(t)\sgn a_{jk}(t)-x_j(t))\,\forall
j. \ee Here $A(t)=(a_{jk}(t))$ is a locally bounded matrix-valued
function which describes the interaction topology of the network and
$a_{jj}(t)\equiv 0$. At time $t\ge 0$, the opinion of the $j$th
agent is influenced by agents for which $a_{jk}\ne 0$
(``neighbors''). Unlike conventional consensus protocols
\cite{Murray:04} this influence may be either cooperative (when
$a_{jk}>0$) or competitive (when $a_{jk}<0$). The coupling term
$|a_{jk}|(x_k\sgn a_{jk}-x_j)$ in \eqref{eq.proto} drives the
opinion of the $j$th agent, respectively, either towards the opinion
of the $k$th one or against it.

In \cite{Altafini:2013} protocol \eqref{eq.proto0} has been
carefully examined, assuming the interaction graph is constant
($A(t)\equiv A$) and strongly connected. It was shown that the
steady-state opinions always agree in modulus, but generally differ
in signs; in other words, the \emph{modulus consensus} of opinions
\cite{MengCaoJohansson:2014} is established.
\begin{defn}
The protocol \eqref{eq.proto0} establishes \emph{modulus consensus}, if for any $x(0)$ a number $x_*\ge 0$ exists such that
\be\label{eq.modulus}
\lim_{t\to+\infty} |x_i(t)|=x_*.
\ee
\end{defn}

The following lemma shows that there are two essentially different types of modulus consensus: ``trivial'' with $x_*=0$ for all $x(0)$ (the system \eqref{eq.proto0} is asymptotically stable)
and ``non-trivial'', where $x_*\ne 0$ for a.a. $x(0)$.
\begin{lemma}\label{lem.struct}
Suppose that protocol \eqref{eq.proto0} establishes modulus
consensus. Then there exist vectors $v,\rho\in\R^N$ with
$\rho_1,\ldots,\rho_N=\pm 1$  such that for any solution of
\eqref{eq.proto0} one has \be\label{eq.v} \lim_{t\to+\infty}
x(t)=\rho v^Tx(0)\Leftrightarrow \lim_{t\to+\infty}
x_j(t)=\rho_jv^Tx(0). \ee
\end{lemma}
Lemma~\ref{lem.struct} shows that in the ``non-trivial'' case $v\ne 0$,
opinions either reach consensus ($\rho_1=\ldots=\rho_N$) or polarize
($\rho_i$ have different signs) whenever $v^Tx(0)\ne 0$. For both
situations we say that the protocol establishes \emph{bipartite
consensus}.
\begin{defn}
We call the protocol \eqref{eq.proto0} \emph{stabilizing}, if $\lim\limits_{t\to\infty} x_j(t)=0\,\forall j\forall x(0)$. The protocol establishes \emph{bipartite consensus} if \eqref{eq.v}
holds with some $v\ne 0$; it establishes \emph{consensus} if additionally $\rho=\bar 1_N$ or $\rho=-\bar 1_N$.
\end{defn}

It was proved in \cite{Altafini:2013} if $A(t)\equiv A$, then the
protocol is  stabilizing (that is, $-L[A]$ is a Hurwitz matrix)
unless the graph $G[A]$ is structurally balanced (SB). The latter
property implies that a community is divided into two hostile camps
(such as votaries of two political parties), where each agent
cooperates with its camp-mates, competing with agents from the
opposite camp. A special case of SB graphs is the graph with
non-negative weights $a_{jk}\ge 0$ where one of the camps is empty.
In this case strong connectivity (SC) and even quasi-strong
connectivity (QSC) imply consensus \cite{RenBeardBook}. The case of
general SB and SC graph is reducible to this case by
means of the \emph{gauge transformation} \cite{Altafini:2013}, which
allows to prove \eqref{eq.modulus}, where $x_*$ depends on the
initial conditions. If both hostile camps are not empty, the
opinions polarize. In other words, structural balance implies
bipartite consensus. A generalization of the gauge transformation
from \cite{Altafini:2013} is the \emph{lifting} approach from
\cite{Hendrickx:14}, splitting each agent into a pair of virtual
agents with opposite opinions, after which the original dynamics can
be considered as a projection of some larger network with purely
cooperative interactions. This approach can be applied also to some
time-varying networks.

In Section~\ref{subsec.static} we further refine Altafini's results
on modulus consensus over static graphs by discarding the strong
connectivity assumption. We show that for structurally balanced
graphs modulus consensus is established if and only if the graph is
QSC; in fact, in this case bipartite consensus is established.
Conversely, bipartite consensus is reached only when the graph is
structurally balanced and QSC. If the graph has no structurally
balanced in-isolated subgraphs, the protocol is stabilizing. Thus we
offer \emph{necessary and sufficient} conditions for modulus
consensus for \emph{general} static graph.

The main concern of this paper is modulus consensus over
\emph{time-varying} signed graphs. In \cite{Altafini:2013} this
problem was considered only for the very special case where the
graph is constantly strongly connected, has time-invariant signs of
the arcs and also \emph{weight-balanced} (this assumption was not
explicitly mentioned, but in fact was used in the proof which
appeals to \cite[Theorem 9]{Murray:04}). Below we relax these
restrictions. Dealing with real-world social networks, the
time-invariance of such relationships between individuals as
friendship and hostility is evidently a non-realistic assumption.
What is more important, the opinion dynamics in social networks are
usually considered to be nonlinear \cite{Krause:2002,Flache:2011}.
Such models are often reducible to the linear case by introducing
time-varying gains, depending on the solution; however, the
corresponding graphs can hardly be weight-balanced. Our techniques
allow us to examine both linear and nonlinear consensus protocols
from \cite{Altafini:2013, Altafini:2012} in the common framework.
Although it is a hard problem to find explicitly the ultimate
opinion vector in the case of time-varying topologies,
Lemma~\ref{lem.struct} shows that there are similarities with the
static case.

A common techniques used to prove consensus in the  case of
cooperative agents is the \emph{shrinking} property of the convex
hull, spanned by agent's opinions. Under the UQSC property of the
graph, the diameter of this convex hull may serve as a Lyapunov
function \cite{Moro:05,Muenz:11,LinFrancis:07}. The UQSC condition
is not necessary in general \cite{Moro:05},  considered as ``the
weakest assumption on the graph connectivity such that consensus is
guaranteed for arbitrary initial conditions'' \cite{Muenz:11}, and
becomes necessary under additional restriction of uniform
convergence \cite{LinFrancis:07}. On the other hand, the EQSC
condition is always necessary for consensus yet insufficient in the
case of directed topologies \cite{Moro:05}. This gap between
necessary and sufficient conditions has been filled recently for
type-symmetric and other cut-balanced graphs
\cite{TsiTsi:13,MatvPro:2013} where EQSC is not only necessary but
also sufficient for consensus.

Under antagonistic interactions between the agents, the convex hull spanned by opinions is not shrinking, and the only available Lyapunov function is the maximal modulus, which will be shown to be non-increasing and thus converging to a limit. However, the UQSC property in general does not guarantee that the minimal modulus also converges to the same limit (as will be shown by a counter example). To provide this, one requires stronger USC conditions. Whereas EQSC property is necessary for bipartite consensus, it is not necessary for stability, as illustrated by the following trivial example. Let $A=\diag(A_1,A_2)$, where both graphs $G[A_1]$ and $G[A_2]$ are strongly connected and structurally unbalanced. As follows from \cite{Altafini:2013}, the matrices $(-L[A_1])$ and $(-L[A_2])$ are Hurwitz, which also holds for $(-L[A])=-\diag(L_1[A],L_2[A])$ and thus the protocol is stabilizing. The static graph $G[A]$ is not QSC and thus not EQSC. Filling this gap between the necessary and sufficient conditions for modulus consensus that is even ``wider'' than in the cooperative case, is a tough open problem. However, adopting the techniques from \cite{MatvPro:2013}, we fill this gap for \emph{cut-balanced} graphs by offering necessary and sufficient conditions of modulus consensus (Subsection~\ref{subsec.cut-balance}).


\section{Main Results}\label{sec.main}

This section is organized as follows. We start with modulus
consensus criteria for static graphs which extend results from
\cite{Altafini:2013} by discarding the assumption of strong
connectivity in them (Subsection~\ref{subsec.static}). We show that
the necessary and sufficient condition for bipartite consensus is
structural balance and QSC, and give also necessary and sufficient
conditions for stability. The next Subsection~\ref{subsec.dir} deals
with the case of general time-varying graphs. We show sufficiency of
the USC condition for modulus consensus and demonstrate that, unlike
the cooperative case, this condition cannot be relaxed to UQSC. In
the last Subsection~\ref{subsec.cut-balance} we focus on modulus
consensus over \emph{cut-balanced} graphs. In this case it is
possible to give necessary and sufficient conditions for both types
of modulus consensus, whereas filling the gap between necessary and
sufficient conditions for modulus consensus under general directed
graphs remains a tough open question.

\subsection{Time-invariant protocols}\label{subsec.static}

Throughout this section $A(t)\equiv A$ is a constant matrix. We
start with the case of structurally balanced graph. In this case, a
gauge transformation \cite{Altafini:2013} exists which reduces  the
protocol to a cooperative one, whose properties are well
established.
\begin{lemma}\label{lem.static.sb}
Let $G[A]$ be structurally balanced. Then $L[A]$ has eigenvalue at $0$ and the following claims are equivalent:
\begin{enumerate}
\item the graph $G[A]$ is QSC;
\item the linear subspace $\ker L[A]\subset\R^n$ has dimension $1$;
\item the protocol \eqref{eq.proto0} establishes modulus consensus.
\end{enumerate}
If these claims hold, then $\rho$, $v$ from Lemma~\ref{lem.struct} are respectively the right and the left eigenvectors of $L[A]$ at $0$, hence $v^TL[A]=L[A]\rho=0$ and $v\ne 0$, so protocol establishes bipartite consensus. If $A$ is a non-negative matrix (one of the hostile camps is empty), the protocol establishes consensus, and otherwise, opinions polarize.
\end{lemma}

In the case where $G[A]$ is structurally balanced yet not QSC  (so
modulus consensus is not reached), the structure of steady-state
opinions may be described in terms of the maximum out-forest matrix
as done in \cite{AgaevChe:2014} for  cooperative agents.

As follows from Lemma~\ref{lem.static.sb}, a protocol with
structurally balanced graph cannot be stabilizing. Therefore,
stability is also impossible when the graph contains an in-isolated
structurally balanced (ISB) subgraph, in other words, the group
involves hostile camps whose members ignore the opinions of the
remaining agents. The nodes of such a subgraph, if existed, would reach
bipartite consensus of opinions independently of the remaining
agents. The following theorem gives a criterion for modulus
consensus over static graphs, showing that existence of ISB
subgraphs is the only obstacle for stability.
\begin{theorem}\label{thm.static}
Let $A(t)=const$. The protocol \eqref{eq.proto0} is stabilizing if and only if the graph is neither SB itself nor contains an ISB subgraph. Bipartite consensus is established if and only if $G[A]$ is structurally balanced and QSC: if $a_{jk}\ge 0\,\forall j,k$, then consensus is established; otherwise, opinions polarize.
\end{theorem}

\begin{figure}
\centering
\includegraphics[width=0.33\columnwidth]{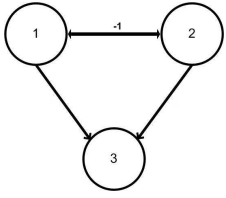}
\caption{Static QSC graph: modulus consensus is impossible if $a_{31}a_{32}>0$}
\label{fig.1}
\end{figure}

\emph{Example 1.} Consider a team of $N=3$ agents with states $x_1(t),x_2(t),x_3(t)$. Assume that $a_{12}=a_{21}=-1$ (see Fig.~\ref{fig.1}) and $a_{31}a_{32}>0$. Thus the equations are
\ben
\begin{split}
\dot x_1=(-x_2-x_1),\,\dot x_2=(-x_1-x_2),\\
\dot x_3=a_{31}x_1+a_{32}x_2-(|a_{31}|+|a_{32}|)x_3,
\end{split}
\een and the graph is not structurally balanced (agents $1$ and $2$
are constantly antagonistic, so the structural balance requires
agent $3$ to cooperate with only one of them, competing with the
other, whereas in reality it cooperates with both  agents 1 and 2).
According to Theorem~\ref{thm.static}, modulus consensus is
impossible (since the ISB subgraph with the set of nodes $\{1,2\}$
exists). This can also be shown in a straightforward way: the system
has equilibria $(\xi,-\xi,\rho\xi)$, with $\rho\dfb
(a_{31}-a_{32})/(|a_{31}|+|a_{32}|)\in (-1;1)$, $\xi\in\R$.

This simple example illustrates that, unlike the cooperative case
($a_{jk}\ge 0$), the protocol with static QSC graphs in general
\emph{does not} establish modulus consensus. To guarantee modulus
consensus, one typically requires \emph{strong} connectivity
(assumed in \cite{Altafini:2013}) or some other property, excluding
ISB subgraphs.


\subsection{Protocols over dynamic signed graphs}\label{subsec.dir}

We start with the following useful lemma, which does not rely on any connectivity assumptions and shows, in particular, that solutions to \eqref{eq.proto0} are always bounded.
\begin{lemma}\label{lem.bound}
For any solution of system \eqref{eq.proto0}, the function $|x(t)|_{\infty}=\max_i|x_i(t)|$ is monotonically non-increasing: $|x(t)|_{\infty}\le |x(t_0)|_{\infty}$ whenever $t\ge t_0\ge 0$. Equivalently, the Cauchy evolutionary matrix $\Phi(t;t_0)$ of system \eqref{eq.proto0} satisfies the inequality $|\Phi(t;t_0)|_{\infty}\le 1$ for $t\ge t_0$.
\end{lemma}

Lemma~\ref{lem.bound} implies, in particular, the existence of the limit $\lim\limits_{t\to+\infty} |x(t)|_{\infty}$. The following theorem shows that under the uniform \emph{strong} connectivity property and bounded coupling gains the modules of all opinions converge to the same limit.
\begin{theorem}\label{thm.usc}
If $a_{jk}(t)$ are bounded and the graph $G[A(\cdot)]$ is USC, then the protocol \eqref{eq.proto0} establishes modulus consensus.
\end{theorem}

As a corollary, we immediately obtain the well-known criterion for consensus under cooperative protocols.
\begin{corollary}\label{cor.conse}
If $a_{jk}(t)\ge 0\,\forall j,k$ for a.a. $t\ge 0$ and $G[A(\cdot)]$ is USC,
then the protocol \eqref{eq.proto0} establishes consensus.
\end{corollary}

It is well known that assumptions of Corollary~\ref{cor.conse} can be relaxed: in the case where $a_{jk}\ge 0$ the UQSC condition is sufficient for consensus \cite{Moro:04,LinFrancis:07,Muenz:11}.
Moreover, using the gauge transformation approach from \cite{Altafini:2013}, sufficiency of the UQSC property may be proved for a dynamic structurally balanced graph, provided that the subdivision into two ``hostile camps'' remains unchanged.
\begin{lemma}\label{lem.fix_division}
Suppose that $V=1:N=V_1\cup V_2$, where $a_{jk}(t)\ge 0$ for any $t\ge 0$ if $j,k\in V_1$ or $j,k\in V_2$; otherwise, $a_{jk}(t)\le 0$ for any $t\ge 0$. If the graph $G[A(\cdot)]$ is UQSC, the protocol establishes bipartite consensus (if $V_1=\emptyset$ or $V_2=\emptyset$) or bipartite consensus (when $V_1,V_2\ne\emptyset$).
\end{lemma}
\begin{remark}
Lemma~\ref{lem.fix_division} obviously remains valid if two hostile
camps $V_1,V_2$  exist only for $t\ge t_0$, where $t_0\ge 0$. This
observation makes the result of Lemma~\ref{lem.fix_division}
applicable to topologies that evolve in order to achieve the
structural balance in finite time (after which the signs of arcs
remain unchanged); graph dynamics leading to structural balance in
finite time were proposed in \cite{Marvel:2011,Antal:2005}.
\end{remark}

However, in general the USC condition in Theorem~\ref{thm.usc} is
not relaxable to UQSC.  Example~1 in Subsection~\ref{subsec.static}
shows that even for static graphs the QSC property (equivalent to
UQSC) does not guarantee modulus consensus unless the graph is
structurally balanced. Our next example shows that the UQSC property
is  not sufficient neither when the graph remains structurally
balanced but the relations of friendship and hostility between the
agents evolve over time. We construct a protocol \eqref{eq.proto0}
with periodic piecewise-constant matrix $A(t)$, such that the graph
$G[A(t)]$ is structurally balanced for any $t\ge 0$ and UQSC, and
nevertheless modulus consensus is not established.

\emph{Example 2.} Consider the more general system
\be\label{eq.aux}
\begin{split}
\dot x_1(t)&=(-x_2(t)-x_1(t)),\,\dot x_2(t)=(-x_1(t)-x_2(t)),\\
\dot x_3(t)&=a_{31}(t)(x_1(t)-x_3(t))+a_{32}(t)(x_2(t)-x_3(t)).
\end{split}
\ee
The functions $a_{31},a_{32}$ are constructed as follows. Consider first system \eqref{eq.aux} with $a_{31}(t)\equiv 1,a_{32}(t)\equiv 0$
and the solution to \eqref{eq.aux} launched at the initial state $x_1(0)=1,x_2(0)=-1,x_3(0)=-1/2$. It is evident that $x_1(t)=1=-x_2(t)$ for any $t\ge 0$ and $x_3(t)\uparrow 1$ as $t\to+\infty$. Therefore, there exists the first time instant $T_0>0$ such that $x_3(T_0)=1/2$. Notice that in the symmetric situation where $a_{31}(t)\equiv 0,a_{32}(t)\equiv 1$ and $x(t)$ is a solution to \eqref{eq.aux} starting at $x_1(0)=1,x_2(0)=-1,x_3(0)=1/2$, one has $x_3(t)\downarrow -1$ and $T_0$ is the first instant where $x_3(T_0)=-1/2$. Taking
$$
a_{31}(t)=1-a_{32}(t)=\begin{cases}
1,\,t\in [0;T_0)\cup [2T_0;3T_0)\cup\ldots\\
0,\,t\in [T_0;2T_0)\cup [3T_0;4T_0)\cup\ldots
\end{cases},
$$
one finally gets a $2T_0$-periodic matrix $A(t)$, corresponding to
the UQSC graph $G[A(\cdot)]$. Moreover, this graph is also
quasi-strongly connected and structurally balanced at any time. Even
so the solution to \eqref{eq.aux} starting at
$x_1(0)=1,x_2(0)=-1,x_3(0)=-1/2$ does not achieve modulus consensus.
It can be easily shown that $x_1(t)=-x_2(t)=1$ for any $t\ge 0$.
Since $a_{31}(t)=1$ and $a_{32}(t)=0$ when $t< T_0$, one has
$x_3(T_0)=1/2$ by definition of $T_0$. On the next interval $t\in
[T_0;2T_0)$ one has $a_{31}(t)=0$ and $a_{32}(t)=1$ and hence
$x_3(2T_0)=-1/2$, so the solution $x(t)$ is periodic and $x_3(t)\in
[-1/2;1/2]$ whereas $|x_1(t)|=|x_2(t)|=1$.

Dealing with purely cooperative protocols, UQSC is considered to be not only sufficient for consensus but also ``nearly'' necessary. It may also be relaxed to EQSC for
 some types of graphs (e.g. cut-balanced ones). On the other hand, the EQSC condition is always necessary for consensus among cooperating agents, being in general not sufficient. As discussed in Section~\ref{subsec.static}, stability of \eqref{eq.proto0} is possible without the EQSC property even for static graphs. However, EQSC is required for bipartite modulus consensus.
\begin{lemma}\label{lem.eqsc}
If the protocol \eqref{eq.proto0} establishes bipartite consensus, the graph $G[A(\cdot)]$ is EQSC.
\end{lemma}

\begin{remark}
In the general situation, where the graph is USC but the assumptions
of Lemma~\ref{lem.fix_division} do not hold, it is difficult to
distinguish between stability and bipartite consensus. A sufficient
condition for stability was proved in \cite{Hendrickx:14}: the
protocol is stabilizing, if for some $\ve>0,T>0$ all the graphs
$G[\int_t^{t+T}\abs A(s)\,ds]$ (where $t\ge 0$) have strongly
connected and structurally unbalanced $\ve$-skeletons
\cite{Hendrickx:14}. This is the only case where the ``lifted''
network is proved to inherit the USC property \cite{Hendrickx:14}.
Generally, this is not the case, so the approach \cite{Hendrickx:14}
does not allow to derive  Theorem~\ref{thm.usc}.
\end{remark}

\subsection{Modulus consensus over cut-balanced graphs}\label{subsec.cut-balance}

In the previous subsection, we get a sufficient condition for modulus consensus (the USC condition). In general, this property is not necessary for modulus consensus. Moreover, even the weaker EQSC condition is necessary for bipartite consensus but not for stability. Filling this gap between necessary and sufficient conditions is a hard open problem even for cooperative protocols. However,
this gap has been recently filled for cooperative protocols with type-symmetric \cite{MatvPro:2013} and more general cut-balanced graphs \cite{TsiTsi:13}. The proofs from \cite{TsiTsi:13} are not applicable for signed graphs. However, adopting the approach from \cite{MatvPro:2013}, we extend the result from \cite{TsiTsi:13} to the case of modulus consensus over signed graph, giving necessary and sufficient conditions for each type of modulus consensus.

Throughout this section the graph $G[A(\cdot)]$ is cut-balanced,
i.e. the inequalities \eqref{eq.cut} hold for some $K\ge 1$. Recall
the $j$th agent essentially interacts with the $k$th one if
$\int_0^{\infty}a_{jk}(t)dt=\infty$ and hence either
$\int_0^{\infty}a_{jk}^+(t)dt=\infty$ or
$\int_0^{\infty}a_{jk}^-(t)dt=\infty$. We say agents
\emph{essentially cooperate} in the first case and \emph{essentially
compete} in the second situation (in general, both relations may
hold). Let $\E\dfb\E[A(\cdot)]$ and $\E^+\subseteq \E, \E^-\subseteq
\E$ be the sets of those pairs of agents $(j,k)$ that respectively
essentially cooperate and essentially compete.

If $\E^+\cap \E^-=\emptyset$, we assign the weights $+1$ and $-1$ to the arcs from $\E^+$ and $\E^{-}$ respectively, transforming $\G\dfb\G[A(\cdot)]$ to a signed graph $\G^{\pm}=(1:N,\E,(s_{jk}))$, $s_{jk}=+1$ for $(j,k)\in \E^+$ and $s_{jk}=-1$ for $(j,k)\in \E^-$. 

As implied by \cite[Lemma 1]{TsiTsi:13}, for cut-balanced graphs the
EQSC property implies ESC, precisely, any quasi-strongly connected
component of the digraph $\G\dfb\G[A(\cdot)]$ is strongly connected,
and a path between $j$ and $k$ exists if and only if a path from $k$
to $j$ exists. From Lemma~\ref{lem.eqsc}, the ESC condition is
necessary for bipartite consensus. In the case of cooperative
agents, this property is also sufficient for  consensus
\cite{TsiTsi:13}. However, in the case of signed graph, ESC is
\emph{not sufficient} without the ``essential'' structural balance.
\begin{theorem}\label{thm.nontrivial}
Assume the graph $G[A(\cdot)]$ is cut-balanced.
The protocol \eqref{eq.proto0} establishes bipartite modulus consensus if and only if $\G^{\pm}$ is well-defined ($\E^+\cap\E^-=\emptyset$), strongly connected and structurally balanced; opinions polarize if and only if $\E^-\ne\emptyset$, and otherwise consensus is established. If $\G$ is strongly connected but $\E^+\cap\E^-\ne\emptyset$ or $\G^{\pm}$ is structurally unbalanced, the protocol \eqref{eq.proto0} is stabilizing.
\end{theorem}

In the case of purely cooperative protocol ($a_{jk}(t)\ge 0$),
Theorem~\ref{thm.nontrivial} transforms into the result obtained in
\cite{TsiTsi:13,MatvPro:2013}.
\begin{corollary}\label{cor.consensus.EQSC}
Cooperative protocol establishes consensus if and only if the graph $G[A(\cdot)]$ is essentially connected.
\end{corollary}

Our next result addresses the case where  $\G$ is not necessarily
connected and thus may be decomposed  into several disjoint strongly
connected components $\G=\G_1\cup \G_2\cup\ldots\cup\G_d$,
$\G_r=(V_r,\E_r)$, $d\ge 1$. In this case in any component $\G_r$,
modulus consensus is established, and the type of which depends only
on the structure of $\G_r$. Let $\E_r^+:=\E_r\cap\E^+$ and
$\E_r^-:=\E_r\cap\E^-$. If $\E_r^+\cap\E_r^-=\emptyset$, define a
signed graph $\G_r^{\pm}$ by assigning arcs from $\E_r^+,\E_r^-$
with weights $+1$ and $-1$ respectively.
\begin{theorem}\label{thm.disconn}
For any solution of \eqref{eq.proto} there exists limits
$x_i^{\dagger}=\lim\limits_{t\to\infty}x_i(t)$ and
$|x_i^{\dagger}|=|x_j^{\dagger}|$ whenever $i$ and $j$ are in the
same strongly connected component: $i,j\in V_r$. If
$\E_r^+\cap\E_r^-=\emptyset$ and $\G_r^{\pm}$ is structurally
balanced, the bipartite consensus is achieved, which comes to
consensus if $\E_r^-=\emptyset$,  and otherwise the opinions
polarize ($V_r=V_r^1\cap V_r^2$ and
$x_i^{\dagger}=-x_j^{\dagger}\,\forall i\in V_r^1,j\in V_r^2$). If
$\E_r^+\cap\E_r^-\ne\emptyset$ or the graph $\G_r^{\pm}$ is
structurally unbalanced, $x_i^{\dagger}=0\,\forall i\in V_r$ that
is, dynamics of opinions from $V_r$ are stable.
\end{theorem}

The following criterion of stability is immediate.
\begin{corollary}\label{cor.trivial}
The protocol \eqref{eq.proto0} is stable if and only if for any
strongly connected component $\G_r$, one has either
$\E_r^+\cap\E_r^-\ne\emptyset$ or $\G_r^{\pm}$ being structurally
unbalanced.
\end{corollary}

\begin{remark}
Theorems~\ref{thm.nontrivial} and \ref{thm.disconn} were proved in \cite{Hendrickx:14} in the special case of ``type-symmetric'' graphs, such that \eqref{eq.unibidir} holds for some $K\ge 1$.
The main idea of the proof is to show that the latter property remains valid for the ``lifted'' network, which is purely cooperative and hence can be examined by techniques from \cite{TsiTsi:13}.
We extend the result from \cite{Hendrickx:14} to cut-balanced graphs. Although this extension seems to be provable by techniques from \cite{Hendrickx:14}, our proof based on \cite{MatvPro:2013} is of independent interest; we elaborate mathematical techniques to cope with both general and cut-balanced cases in similar ways.
\end{remark}


\section{Applications: Nonlinear Protocols}\label{sec.apply}

In this section we apply our results to some types of nonlinear consensus protocols, similar to those from \cite{Altafini:2012,Altafini:2013}.

\subsection{Additive Laplacian protocols}

Our first example concerns \emph{nonlinear} consensus algorithms that are referred in \cite{Altafini:2013} as the ``additive Laplacian feedback schemes''. The first of them is
\be\label{eq.nonlin2}
\dot x_i(t)=\sum_{j=1}^N|a_{ij}(t)|(h_{ij}(x_j(t)\sgn a_{ij}(t))-h_{ij}(x_i(t))),
\ee
and the second protocol has the form
\be\label{eq.nonlin3}
\dot x_i(t)=\sum_{j=1}^N|a_{ij}(t)|h_{ij}(x_j(t)\sgn a_{ij}(t)-x_i(t)))\,\forall i.
\ee

We adopt the following assumption about the nonlinearities.
\begin{assum}\label{ass.h}
For any $i,j\in 1:N$ the map $h_{ij}\in C^1(\R)$ is strictly
increasing (and hence $h_{ij}'>0$) with $h_{ij}(0)=0$.
\end{assum}

Defining the functions $H_{ij}[y,z]$ as follows:
$H_{ij}[y,z]:=(h_{ij}(y)-h_{ij}(z))/(y-z)$ for $y\ne z$ and
$H_{ij}[z,z]:=h_{ij}'(z)$ so that $H_{ij}[y,z]>0$ for any $y,z$.
Since $h_{ij}\in C^1$, $H_{ij}$ is easily shown to be a continuous
function and $h_{ij}(y)-h_{ij}(z)=H_{ij}[y,z](y-z)\,\forall y,z$.
Under Assumption~\ref{ass.h},
Theorems~\ref{thm.usc} and \ref{thm.nontrivial} appears to be applicable
to the protocols \eqref{eq.nonlin2}, \eqref{eq.nonlin3} after the
standard trick, replacing nonlinearities with the solution-dependent
gains, as shown by the following lemma.
\begin{lemma}\label{lem.tech-nonlin}
Let $x(t)$ be a solution to system \eqref{eq.nonlin2}, which is defined for $t\ge 0$. Define the matrix $\mathfrak A(t)=(\a_{ij}(t))$ by $\a_{ij}(t):=a_{ij}(t)H_{ij}[x_j(t)\sgn a_{ij}(t),x_i(t)]$. Then
\be\label{eq.tech2}
\dot x(t)=-L[\mathfrak A(t)]x(t).
\ee
If the graph $G[A(\cdot)]$ is EQSC, ESC, UQSC, USC, or cut-balanced, then the same is valid for the graph $G[\mathfrak A(\cdot)]$.
If the matrix $A(\cdot)$ is globally bounded, the same is valid for the matrix $\mathfrak A(\cdot)$.
These claims also hold for the protocol \eqref{eq.nonlin3}, taking $\a_{ij}(t):=a_{ij}(t)H_{ij}[x_j(t)\sgn a_{ij}(t)-x_i(t),0]$.
\end{lemma}

Application of Theorems~\ref{thm.usc}, \ref{thm.nontrivial}  to \eqref{eq.tech2} yields the following.
\begin{theorem}\label{thm.nonlin2}
Under Assumption~\ref{ass.h}, the solutions to systems \eqref{eq.nonlin2}, \eqref{eq.nonlin3} exist, are unique and infinitely prolongable for any initial condition.
If the graph $G[A(\cdot)]$ is USC and $A(\cdot)$ is bounded, or the graph $G[A(\cdot)]$ is ESC and cut-balanced, the protocols \eqref{eq.nonlin2}, \eqref{eq.nonlin3} establish modulus consensus.
\end{theorem}

Comparing the result of Theorem~\ref{thm.nonlin2} with that of \cite[Theorem 3,4]{Altafini:2013}, one notices that our assumption about the nonlinearities $h_{ij}$ differs from \cite{Altafini:2013}, where they are not assumed to be smooth, but only monotonic with some integral constraint. However, unlike \cite{Altafini:2013}, functions $h_{ij}$ may be heterogeneous and not necessarily odd; the graph may be time-varying.

Note that Theorem~\ref{thm.nonlin2} gives only sufficient conditions
for modulus consensus. Necessary  conditions such as
Lemma~\ref{lem.eqsc} are not directly applicable since they assume
the matrix $A(t)$ to be common for all solutions. Extending the
concept of essentially equivalent protocols (see
Subsection~\ref{subsec.proof.3}) to the nonlinear case, it is
possible to get some necessary conditions which are, however, beyond
the scope of this paper.

\subsection{Nonlinear Laplacian Flow}

In this subsection we examine the nonlinear consensus protocol
similar to that addressed in \cite[Section IV-B]{Altafini:2013}:
\be\label{eq.nonlin1} \dot
x_i(t)=\sum_{j=1}^N|F_{ij}(t,x)|(x_j(t)\sgn F_{ij}(t,x)-x_i(t)), \ee
here $i\in 1:N$ and $F_{ij}:[0;\infty)\times\R^N\to\R$ are
Caratheodory maps, i.e. $F_{ij}(t,\cdot)$ are continuous for a.a.
$t$ and $F_{ij}(\cdot,x)$ are measurable for any $x$. We assume also
that for any compact set $K\subset [0;\infty)\times\R^N$ one has
\be\label{eq.auxbound} \sup\{F_{ij}(t,x):x\in K,t\ge
0\}<\infty\quad\forall i,j. \ee
\begin{theorem}\label{thm.nonlin1}
For any initial condition $x(0)$ a solution of \eqref{eq.nonlin1} exists for $t\ge 0$ and the matrix $A(t)\dfb F_{ij}(t,x(t))$ is bounded. If the graph $G[A(\cdot)]$
is USC or ESC and cut-balanced, the protocol \eqref{eq.nonlin1} establishes modulus consensus.
\end{theorem}

Although in general it is hard to verify the uniform or essential
strong connectivity of $G[A(\cdot)]$, where $A(t)=F_{ij}(t,x(t))$
depends on the concrete solution, in special cases such a property
may be proved. For instance, it is implied by the \emph{global
strong $\ve$-connectivity} \cite[Section IV-B]{Altafini:2013}: the
graph $G(\hat F_{ij}(t,x))$ is strongly $\ve$-connected for any
$t,x$. The result of Theorem~\ref{thm.nonlin1} extends the result
from \cite[Section IV-B]{Altafini:2013} in several ways. First of
all, it deals with time-variant gains $F_{ij}(t,x)$ and does not
require them to have a constant sign. In particular, system
\eqref{eq.nonlin1} does not necessarily generate order-preserving
flow \cite{Altafini:2012}. Moreover, we do not assume that the graph
$G[A(t)]$ is weight-balanced which can hardly be provided for
nonlinear functions $F_{ij}$. In the case of USC graphs, the balance
is not needed at all; in the case of ESC graphs it is replaced by
the much weaker cut-balance condition. At last, we relax the
connectivity assumption.


\section{Proofs}\label{sec.proof}

We start with the proof of Lemma~\ref{lem.struct} and then prove results, concerned with static graphs (Subsection~\ref{subsec.proof.2}).
To proceed with the case of dynamic graph, we elaborate some useful techniques in Subsections~\ref{subsec.order} and \ref{subsec.proof.3}, 
entailing also Lemmas~\ref{lem.bound} and \ref{lem.eqsc}. The case of cut-balanced graph is considered in Subsection~\ref{subsec.proof.4}.
In Subsections~\ref{subsec.usc} and \ref{subsec.nlin} we prove the modulus consensus criterion for directed dynamic graphs and its implications, dealing with nonlinear protocols.

Throughout the section, $\Phi(t|t_0)$ (where $t,t_0\ge 0$) stands for the Cauchy evolutionary matrix of the system \eqref{eq.proto0}, that is,
the solution of the Cauchy problem for \eqref{eq.proto0} with initial data $x(t_0)=x_0$ is given by
$x(t)=x(t|t_0,x_0)=\Phi(t|t_0)x(0)$.

\begin{IEEEproof}[Proof of Lemma~\ref{lem.struct}]
 Assume the protocol \eqref{eq.proto0} establishes modulus consensus. Note that since functions $x_i(t)$ are continuous, existence of the limits $\lim\limits_{t\to+\infty}|x_i(t)|=x_*$ implies that the limits $\lim\limits_{t\to+\infty}x_i(t)$ also exist (and equal to $\pm x_*$). Therefore $\Phi(t)\underset{t\to\infty}{\longrightarrow}\Phi_*:=[\phi_1,\ldots,\phi_N]$ as $t\to\infty$, where each column $\phi_j$ have entries with equal modules. The same applies for any linear combination $\sum_{j=1}^N\alpha_j\phi_j$. If $\Phi_*=0$, the statement of Lemma~\ref{lem.struct} is evident, taking $v=0$. Assume that one of $\phi_j$, say, $\phi_1$ is nonzero, thus $\phi_1=v_1\rho $ where $v_1\ne 0$ and $\rho$ is a vector with entries $\pm 1$. Notice that for any real numbers $\alpha,\beta\ne 0$ we have $|\alpha-\beta|\ne |\alpha+\beta|$.
Therefore, if $\phi_j\ne 0$ for some $j\ne 1$, all entries of $\phi_j-\phi_1$ have the same module if and only if $\phi_j=v_j\rho$, $v_j\ne 0$.
If $\phi_j=0$, we put by definition $v_j=0$. Therefore, $\phi_j=v_j\rho$ for any $j$ and $\lim\limits_{t\to\infty} x_j(t)=\Phi_*=\rho v^Tx(0)$, where $v:=(v_1,\ldots,v_N)^T$.
\end{IEEEproof}

\subsection{Proofs of Lemmas~\ref{lem.static.sb}, \ref{lem.fix_division} and Theorem~\ref{thm.static}}\label{subsec.proof.2}

\begin{IEEEproof}[Proof of Lemma~\ref{lem.static.sb}]
 We use a \emph{gauge transformation} \cite{Altafini:2013}. Suppose that the graph is structurally balanced; let $V_1$ and $V_2$ be hostile camps covering all the nodes. Introducing a diagonal matrix $D=\diag(d_1,\ldots,d_N)$ by $d_i=1$ for $i\in V_1$ and $d_i=-1$ for $i\in V_2$, one easily shows that the gauge transformation $x\mapsto z:=Dx$ transforms the system \eqref{eq.proto0} into
\be\label{eq.aux1}
\dot z(t)=-L[\abs A]z(t).
\ee
The properties of cooperative protocols \eqref{eq.aux1} are widely known \cite{AgaevChe:2005,RenBeardBook,MesbahiEgerBook}.
Since the matrix $B=\abs A$ is non-negative, $L[B]$ has zero eigenvalue with eigenvector $1_N$. The algebraic and geometric multiplicities of this eigenvalue coincide \cite{AgaevChe:2005}, and it is simple if and only the graph $G[B]$ is QSC (has oriented spanning tree). If this holds, the protocol \eqref{eq.aux1} establishes consensus and $\exp(-L[B]t)\underset{t\to+\infty}{\longrightarrow} \bar 1_Nv$. Since $L[B]=DL[A]D$ and $G[A]$ is QSC if and only if $G[B]$ is QSC, one immediately obtains the claims of Lemma~\ref{lem.static.sb}.
\end{IEEEproof}

\begin{IEEEproof}[Proof of Lemma~\ref{lem.fix_division}]
The proof employs the same idea of the gauge transformation, retracing the arguments from \cite[Section III-B-2)]{Altafini:2013}.
Suppose the graph $G[A(t)]$ is structurally balanced for any $t$ with static hostile camps $V_1$ and $V_2$. The gauge transformation $x\mapsto z:=Dx$, introduced in the foregoing,
transforms the system \eqref{eq.proto0} into \eqref{eq.aux1}. Since $G[\abs A(t)]$ is UQSC, the protocol \eqref{eq.aux1} establishes consensus \cite{RenBeardBook},\cite{Moro:04} and thus
opinions in network \eqref{eq.proto0} polarize.
\end{IEEEproof}

To proceed with the proof of Theorem~\ref{thm.static}, note that stability is equivalent to the asymptotic stability of the system \eqref{eq.proto0}, which means that
$(-L[A])$ is a Hurwitz matrix. By using the Gershgorin disk theorem, it was shown in \cite{Altafini:2013} that all eigenvalues of $L[A]$, except for possibly zero, have positive real parts.
Hence protocol is stable if and only if $0$ is not an eigenvalue of $L[A]$, i.e. $L[A]\rho=0$ is impossible when $\rho\ne 0$. We need the following simple lemma.
\begin{lemma}\label{lem.static-technical}
Let $L[A]\rho=0$ and $|\rho|_{\infty}=1$. Then $V_1=\{j:\rho_j=1\}$, $V_2=\{j:\rho_j=-1\}$ are hostile camps and $V_*\dfb V_1\cup V_2\ne\emptyset$.
If $j\in V_*,k\not\in V_*$, then $a_{jk}=0$.
\end{lemma}
\begin{IEEEproof}
By assumption, $|x_k|\le 1$ for any $k$ and $|x_j|=1$ for some $j$, hence $V_*\ne\emptyset$. For any such $j$ one has $\rho_j\sum\limits_{k\ne j}|a_{jk}|=\sum\limits_{k\ne j}\rho_ka_{jk}$. We note that $\left|\sum\limits_{k\ne j}\rho_ka_{jk}\right|\le \sum\limits_{k\ne j}|a_{jk}|$. The equality implies, firstly, that $|\rho_ka_{jk}|=|a_{jk}|$ (i.e. $k\in V_*$ or $a_{jk}=0$) and secondly, all non-zero terms $\rho_ka_{jk}$ have the same sign (coinciding with $\sgn\rho_j$). In other words, if $j\in V_*$, then $a_{jk}\rho_j\rho_k\ge 0\,\forall k\in V_*$, and $a_{jk}=0$ when $k\not\in V_*$, from where the statement is straightforward.
\end{IEEEproof}

\begin{IEEEproof}[Proof of Theorem~\ref{thm.static}]
Sufficiency in the first statement is immediate from Lemma~\ref{lem.static-technical}: if no ``in-isolated'' structurally balanced subgraph exists, then $0$ is not an eigenvalue of $(-L[A])$, which therefore is a Hurwitz matrix. Necessity follows from Lemma~\ref{lem.fix_division}: nodes of the subgraph are independent of the remaining agents and reach bipartite consensus, and hence stability of the whole community is impossible. Lemma~\ref{lem.fix_division} implies sufficiency in the second statement. To prove necessity, suppose that bipartite consensus is established. Then $0$ is an eigenvalue of $L[A]$ and there exists an eigenvector $\rho$, such that $L[A]\rho=0$, and $|\rho|_{\infty}=1$. Bipartite consensus implies that for this vector one has $|\rho_1|=\ldots=|\rho_N|=1$. Applying Lemma~\ref{lem.static-technical}, we obtain the structural balance of $G[A]$ since $V_*=1:N$. 
\end{IEEEproof}

\subsection{Ordering permutations}\label{subsec.order}

In this Subsection, we elaborate some useful techniques to be used in the subsequent proofs.

Given a family of scalar functions $f_1(t),\ldots,f_N(t)$ (where $t\ge 0$), let  $[k^1(t), \ldots, k^N(t)]$ be {\it the ordering permutation}, sorting the set $\{f_1(t),\ldots, f_N(t)\}$ in the ascending order: $f_{k^1(t)}(t)\le f_{k^2(t)}(t)\le\ldots\le f_{k^N(t)}(t)$. If $f_j(t)=f_k(t)$ for some $j$,$k$ and $t$, the permutation is not uniquely defined. The following technical Lemma shows the permutation may always be taken in some regular way.
\begin{lemma}\label{lem.permutation}
Assume that $f_j$ are locally Lipschitz. Then there exists such an ordering permutation $k^1(t),\ldots,k^N(t)$ that $k^j(\cdot)$ are measurable, functions $F_j(t)\dfb f_{k^j(t)}(t)$ are locally Lipschitz and $\dot F_j(t)=\dot f_{k^j(t)}(t)$ for any $j$ and a.a. $t\ge 0$.
\end{lemma}

To prove Lemma~\ref{lem.permutation}, we need the following proposition.
\begin{lemma}\label{lem.lip} Let $f_*(t) :=\max_{i\in[1:N]} f_i(t)$ and $j(t)$ be the index such that $f_*(t) =
y_{j(t)}(t)$ for a.a. $t$. In the case of non-uniqueness, one always may choose $j(\cdot)$ in a way that it is measurable
and $\dot f_*(t) = \dot y_{j(t)}(t)$ for a.a $t$. The claim remains valid, replacing $\max$ with $\min$.
\end{lemma}
\begin{IEEEproof}
As follows from the generalized version of Danskin theorem \cite[Theorem 2.1]{Clarke:1975}, $f_*(\cdot)$ is locally Lipschitz
and $\dot f_*(t)\in\{\dot y_j(t):y_j(t)=f_*(t)\}$ for a.a. $t\ge 0$. The Filippov-Castaing measurable selector theorem
(see e.g. \cite[Theorem 1]{McShane:1967}) yields that a measurable function $j(t)$ exists such that $\dot f_*(t) = \dot y_{j(t)}(t)$ and
$y_{j(t)}(t)=f_*(t)$. The last claim is proved by replacement $f_j\mapsto -f_j$.
\end{IEEEproof}

\begin{IEEEproof}[Proof of Lemma~\ref{lem.permutation}] The proof is via induction on $N$. For $N=1$, the claim is evident. Let
it be true for some $N$, and let $f_1(\cdot), \ldots, f_{N+1}(\cdot)$ be locally Lipschitz.
The ordering of the first $N$ functions $z_1(\cdot), \ldots,z_N(\cdot)$ are locally Lipschitz by the induction hypothesis. The recursion
$y_{N+1}^{0}(t) := y_{N+1}(t)$,
\begin{gather*}
y_{N+1}^{\nu}(t) := \max\{z_\nu(t); y_{N+1}^{\nu-1}(t) \}, \\ \widehat{z}_\nu(t) := \min\{z_\nu(t);
y_{N+1}^{\nu-1}(t) \}, \; \nu\in 1:N
\end{gather*} results in the ordering
$\widehat{z}_1(t) \leq \ldots \leq \widehat{z}_N(t) \leq
\widehat{z}_{N+1}(t) := y_{N+1,N}(t)$ of the entire set $y_1(\cdot), \ldots,
y_{N+1}(\cdot)$. By applying Lemma~\ref{lem.lip} at every recursion step, we
see that $\widehat{z}_\nu(\cdot)$ are locally Lipschitz. For any $\nu=1,\ldots, N$, the sequence
\begin{equation}
\label{seq.hat} \mathscr{Z}_\nu(t) := [\widehat{z}_1(t),\widehat{z}_2(t),
\ldots,\widehat{z}_{\nu}(t), z_{\nu+1}(t), \ldots, z_N(t), y_{N+1}^{\nu}(t)]
\end{equation}
is obtained from $\mathscr{Z}_{\nu-1}$ via a permutation $J_\nu(t)$ of
indices, which either is the identity one or exchanges the places of the
$\nu$th and $(N+1)$th entries and may be chosen measurable. By Lemma~\ref{lem.lip}, the sequence
$\mathscr{Z}_\nu^\prime(t)$ that results from replacement of any function in
\eqref{seq.hat} by its derivative is related to
$\mathscr{Z}_{\nu-1}^\prime(t)$ by the same permutation for a.a. $t$. The
sequence $\mathscr{Z}_0(t)$ is obtained from $\mathscr{Y}(t) :=[y_1(t),
\ldots, y_{N+1}(t)]$ via a permutation of indices $\mathscr{K}(t) = [k^1(t),
\ldots,k^N(t), N+1]$. By the induction hypothesis, $\mathscr{K}(t)$ also
relates $\mathscr{Z}_{0}^\prime(t)$ and
$\mathscr{Y}^\prime(t):=[y_1^\prime(t), \ldots, y_{N+1}^\prime(t)]$ for a.a.
$t$. Then $\mathscr{K}_{N+1}(t)=J_N\circ\cdots\circ J_1\circ \mathscr{K}(t)$
transforms $\mathscr{Y}^\prime(t)$ into $\mathscr{Z}_N^\prime(t)$ for a.a. $t$, which proves the induction step. 
\end{IEEEproof}

\subsection{Some Technical Lemmas and Proofs of Lemmas~\ref{lem.bound},~\ref{lem.eqsc}}\label{subsec.proof.3}

To start with, we note that since $a_{jk}(t)$ are locally bounded, this also holds for $\dot x_j$ and therefore $x_j(\cdot)$ are locally Lipschitz.
For any solution $x(t)$ of \eqref{eq.proto0} we introduce functions
\ben
\chi_j(t):=|x_j(t)|,\,\theta_{ij}(t):= \sgn a_{ij}(t)\sgn x_i(t)\sgn x_j(t).
\een
The following lemma gives a useful interpretation of dynamics \eqref{eq.proto0} in terms of the moduli functions $\chi_k$.
\begin{lemma}\label{lem.newproto}
The functions $\chi_j$ are locally Lipschitz, thus absolutely continuous. For a.a. $t\ge 0$ and any $k$ one has
\be\label{eq.deriv_mod}
\dot\chi_k(t)=\sum\limits_{i=1}^N|a_{ki}(t)|[\chi_i(t)\theta_{ki}(t)-\chi_k(t)].
\ee
(since $\chi_k\ge 0$, it has sign $\sgn \chi_k(t)$ equal to $0$ or $1$).
\end{lemma}
\begin{IEEEproof}
Since $a_{jk}$ are assumed to be locally bounded, $x_k(\cdot)$ are locally Lipschitz. The same applies to $\chi_k(\cdot)$ since
$|\chi_k(t_1)-\chi_k(t_2)|=||x_k(t_1)|-|x_k(t_2)||\le |x_k(t_1)-x_k(t_2)|$. Therefore $\chi_k$ are absolutely continuous and for a.a. $t>0$ the derivative $\dot\chi_k(t)$ exists.
For such $t$ we immediately have $\chi_k(t)=0\Longrightarrow \dot\chi_k(t)=0$ by the Fermat theorem since $0$ is the global minimum of $\chi_k$, which proves \eqref{eq.deriv_mod}
(indeed, $x_k(t)=0$ implies that $\theta_{ki}(t)=0$ for any $i$). Let $\chi_k(t)>0$. Since $\sgn x_k(s)=\sgn x_k(t)$ for $s\approx t$, one has
\ben
\begin{split}
\dot\chi_k(t)&=\dot x_k(t)\sgn x_k(t)\overset{\text{\eqref{eq.proto}}}{=}\\
&\overset{\text{\eqref{eq.proto}}}{=}\sum_{i=1}^N|a_{ki}(t)|[x_i(t)\sgn a_{ki}(t)\sgn x_k(t)-\chi_k(t)],
\end{split}
\een
which proves \eqref{eq.deriv_mod} since $x_i(t)=\chi_i(t)\sgn x_i(t)$.
\end{IEEEproof}

Henceforth, we fix some \emph{ordering permutation $k^1(t),\ldots,k^N(t)$} for the family $\chi_1(t),\ldots,\chi_N(t)$ (see Subsect.\ref{subsec.order}) and put $M_j(t)=\chi_{k^j(t)}$, in particular, $M_N(t)=\max_j|x_j(t)|$.
Combining Lemmas~\ref{lem.newproto} and \ref{lem.permutation}, one gets the following.
\begin{lemma}\label{lem.newprotoM}
The function $M_N(t)=\max\limits_{i\in 1:N}|x_i(t)|$ is non-increasing and hence $\int_0^{\infty}|\dot M_N(t)|dt<\infty$.
For $\tilde a_{ji}:=a_{k^jk^i}$, $\tilde\theta_{ji}:=\theta_{k^jk^i}$, for a.a. $t\ge 0$ and any $j\in 1:N$ one has
\be\label{eq.deriv_mod1}
\dot M_j(t)=\sum\limits_{i=1}^N|\tilde a_{ji}(t)|[M_i(t)\tilde\theta_{ji}(t)-M_j(t)].
\ee
\end{lemma}
\begin{IEEEproof}
Combining \eqref{eq.deriv_mod} with Lemma~\ref{lem.permutation}, we get
\ben
\begin{split}
\dot M_j(t)&=\sum\limits_{i=1}^N|a_{k^ji}(t)|\left[\chi_i(t)\theta_{k^ji}(t)-M_j(t)\right]\\
&=\sum\limits_{i=1}^N|a_{k^jk^i}(t)|\left[\chi_{k^i}(t)\theta_{k^jk^i}(t)-M_j(t)\right],
\end{split}
\een
from where \eqref{eq.deriv_mod1} follows since $M_i=\chi_{k^i}$. Using \eqref{eq.deriv_mod1}, one has $\dot M_N(t)\le 0$ for a.a. $t\ge 0$ since $|\tilde\theta_{ij}|\le 1$ and
$M_j\le M_N$ for any $j$ thus $M_N(\cdot)$ is non-increasing. We have also $\int_0^{\infty}|\dot M_N(t)|dt=M(0)-\inf_{t\ge 0}M(t)\le M(0)$.
\end{IEEEproof}

We also require one additional simple tool which allows to examine the behavior of system \eqref{eq.proto} by comparing it with a simpler system, obtained by ignoring all
inessential interactions between the agents. Consider a protocol
\be\label{eq.proto1}
\dot\xi(t)=-L[\mathfrak A(t)]\xi(t)
\ee
where $\mathfrak A(t)=(\a_{ij}(t))$ is locally bounded. We say the protocol \eqref{eq.proto1} is \emph{essentially equivalent} to \eqref{eq.proto0} if
$$
\int_0^{\infty}|\a_{ij}(t)-a_{ij}(t)|dt<\infty\quad\forall i,j.
$$
We are going to show that the essentially equivalent protocol
provides the same limit sets for the solutions. Let
$B_1:=\{x\in\R^N:|x|_1\le 1\forall i\}$ be a unit ball in the
$|\cdot|_{\infty}$-norm, positively invariant by
Lemma~\ref{lem.newprotoM}: $x(t_0)\in B_1\Longrightarrow x(t)\in
B_1\,\forall t\ge t_0$.
\begin{defn}\label{def.omega}
Let $\Omega_{t_0,x_0}:=\{y\in\R^N:\exists
t_n\to\infty:x(t_n|t_0,x_0)\underset{n\to\infty}{\longrightarrow}
y)\}$. We call the set
$\Omega:=\overline{\bigcup\limits_{t_0,x_0}\Omega_{t_0,x_0}}\subseteq B_1$
\emph{the $\Omega$-set} of the system \eqref{eq.proto0} (the union
is over $t_0\ge 0,x_0\in B_1$).
\end{defn}
\begin{lemma}\label{lem.omega}
Suppose the protocols \eqref{eq.proto0} and \eqref{eq.proto1} are essentially equivalent. Then for any $\ve>0$ there exists $T_0=T_0(\ve)$ such that $\xi(T_0)=x(T_0)\in B\Longrightarrow |x(t)-\xi(t)|_{\infty}\le\ve\,\forall t\ge T_0$. In particular, the systems \eqref{eq.proto0} and \eqref{eq.proto1} have equal $\Omega$-sets.
\end{lemma}
\begin{IEEEproof}
Since $x(t_0)\in B_1\Longrightarrow x(t)\in B_1\forall t\ge t_0$ thanks to Lemma~\ref{lem.newprotoM}, one has $|\Phi(t|t_0)|_{\infty}\le 1\forall t_0\ge 0\forall t\ge t_0$.
Let $T_0>0$ be so large that
$
\int_{T_0}^{\infty}|L[A(t)]-L[\mathfrak A(t)]|_{\infty}dt<\ve.
$
Applying Lemma~\ref{lem.newprotoM} to \eqref{eq.proto1} implies that $\xi(T_0)\in B\Longrightarrow\xi(t)\in B\forall t\ge T_0$, and hence
$\Delta(t):=(L[A(t)]-L[\mathfrak A(t)])\xi(t)$ satisfies the inequality $\int_{T_0}^{\infty}|\Delta(t)|_{\infty}dt<\ve$. Since $\xi'(t)=-L[A(t)]\xi+\Delta(t)$, condition $x(T_0)=\xi(T_0)\in B$ implies
$$
\xi(t)-x(t)=\xi(t)-\Phi(t|T_0)\xi(T_0)=\int_{T_0}^t\Phi(t|s)\Delta(s)ds
$$
and therefore $|\xi(t)-x(t)|_{\infty}\le\int_{T_0}^t|\Phi(t|s)|_{\infty}|\Delta(s)|_{\infty}ds<\ve$.
This proves the first claim from where the second one immediate follows: indeed, for any solution $x(t|t_0,x_0)$ with $t_0\ge 0,x_0\in B$ and any $\ve>0$ one can find $T_0$ such that $|\xi(t|T_0,\xi_0)-x(t|t_0,x_0)|_{\infty}=|\xi(t|T_0,\xi_0)-x(t|T_0,\xi_0)|_{\infty}\le\ve$ for any $t\ge T_0$, where
$\xi_0:=x(T_0|t_0,x_0)\in B$. Therefore, any set $\Omega_{t_0,x_0}$ from Definition~\ref{def.omega} belongs to the $\Omega$-set of \eqref{eq.proto1} and thus the whole $\Omega$-set of \eqref{eq.proto0} belongs to the $\Omega$-set of \eqref{eq.proto1}. The inverse inclusion is proved in the same way.
\end{IEEEproof}
\begin{remark}\label{rem.omega}
Such properties of the protocol as modulus consensus, stability, bipartite consensus, and ``partial'' modulus consensus (modulus agreement among a subgroup of agents) in fact depend only on the $\Omega$-set. For instance, modulus consensus is established if and only if the $\Omega$-set is comprised of the set $\{x\in B:|x_1|=\ldots=|x_N|\}$. Lemma~\ref{lem.omega} implies that those properties are preserved, replacing the protocol with essentially equivalent one.
\end{remark}

We are now going to prove Lemmas~\ref{lem.bound}, \ref{lem.eqsc}. Lemma~\ref{lem.bound} immediately follows from Lemma~\ref{lem.newprotoM} since $M_N(t)=\max_j|x_j(t)|$.

\begin{IEEEproof}[Proof of Lemma~\ref{lem.eqsc}] Suppose the protocol \eqref{eq.proto0} establishes bipartite consensus. Lemma~\ref{lem.omega} and Remark~\ref{rem.omega} show that, without loss of generality, one may assume $a_{jk}\equiv 0$ unless $j$ essentially interacts with $k$, i.e. $(j,k)\in\mathcal E[A(\cdot)]$. If the topology is not EQSC, then the graph $\G[A(\cdot)]$ is not QSC and thus, as shown in \cite{Moro:05}, there exist non-empty disjoint subsets $V_1,V_2\subset 1:N$ that has no incoming arcs: $a_{jk}=0$ if $j\in V_1,k\not\in V_1$ or $j\in V_2,k\not\in V_2$. Therefore, the opinions of agents from $V_1$ are independent on the opinions of agents from $V_2$, and hence bipartite consensus is impossible.
\end{IEEEproof}

\subsection{Proof of Theorems~\ref{thm.nontrivial},~\ref{thm.disconn} and Corollary~\ref{cor.trivial}}\label{subsec.proof.4}

Henceforth $G[A(\cdot)]$ is cut-balanced. The cornerstone of the proofs is the following lemma, based on Lemma~\ref{lem.newprotoM}.
\begin{lemma}\label{lem.main}
For a given solution of \eqref{eq.proto0}, let $\tilde\eta_{ji}(t):=|\tilde a_{ji}(t)|(\tilde\theta_{ji}(t)M_i(t)-M_j(t))$, where $\tilde\theta_{ji}$ are the same as in \eqref{eq.deriv_mod}. Then $\tilde\eta_{ji}\in L^1[0;\infty]$ for any $i,j$, so that $\dot M_j\in L^1[0;\infty]$.
\end{lemma}
\begin{IEEEproof}
Lemma~\ref{lem.newprotoM}, which is valid for any dynamic graph, shows that $\tilde\eta_{N1},\ldots,\tilde\eta_{N,N-1}\in L^1[0;\infty]$ since $\tilde\eta_{Nj}\le 0$ and $\dot M_N\in L^1$.
We are going to show that $\tilde\eta_{N-1,j}\in L^1$.

Note that $|\tilde\eta_{Nj}(t)|\ge |\tilde a_{Nj}(t)|(M_N(t)-M_{N-1}(t))$ for any $j<N$. Indeed, $|\tilde\theta_{N,j}M_j-M_N|\ge |M_N|-|\tilde\theta_{N,j}M_j|\ge M_N-M_{N-1}$ since $M_j\le M_{N-1}$. Applying the definition of cut-balance \eqref{eq.cut} to $V'=\{k^1,k^2,\ldots,k^{N-1}\}$ and $V''=\{k^N\}$, one obtains $\sum\limits_{j<N}|\tilde a_{jN}(t)|\le K\sum\limits_{j<N}|\tilde a_{Nj}(t)|$.
Therefore the function $S_N(t)\dfb (M_N(t)-M_{N-1}(t))\sum\limits_{j=1}^{N-1}|\tilde a_{jN}(t)|$ is summable.
   Note that if $\tilde\eta_{N-1,N}(t)>0$ then $\tilde\theta_{N-1,N}(t)=1$ and $\tilde\eta_{N-1,N}(t)=|a_{N-1,N}(t)|(M_N(t)-M_{N-1}(t))$. Thus, $\tilde\eta_{N-1,N}^+(t)\le S_N(t)$ and hence $\tilde\eta_{N-1,N}^+\in L^1$. By invoking \eqref{eq.deriv_mod1} for $j=N-1$, one obtains that
\be\label{eq.aux2}
\dot M_{N-1}(t)=-\tilde\eta_{N-1,N}^-(t)+\sum_{j=1}^{N-2}\tilde\eta_{N-1,j}(t)+\tilde\eta_{N-1,N}^+(t).
\ee
Since $-\tilde\eta_{N-1,N}^-(t)\le 0$ and $\tilde\eta_{N-1,j}(t)\le 0$ for any $t\ge 0$ and the last term in \eqref{eq.aux2} is $L^1$-summable, we either have $\tilde\eta_{N-1,N}^-,\tilde\eta_{N-1,j}\in L^1$ or $\int_0^{\infty}\dot M_{N-1}(t)dt=-\infty$. The latter is impossible since $M_{N-1}(t)\ge 0$. Thus $\tilde\eta_{N-1,j}\in L^1[0;\infty]\,\forall j$.

Our next step is to prove that  $\tilde\eta_{N-2,j}\in L^1$ for any $j$. We note that for any $j\le N-2$ and $r=N-1,N$ we have  $|\tilde\eta_{rj}(t)|\ge |\tilde a_{rj}(t)|(M_{N-1}(t)-M_{N-2}(t))$.
Applying \eqref{eq.cut} to $V'=\{k^1,k^2,\ldots,k^{N-2}\}$ and $V''=\{k^{N-1},k^N\}$, we obtain that $\sum\limits_{j=1}^{N-2}\sum\limits_{r=N-1}^N|\tilde a_{jr}(t)|\le K\sum\limits_{j=1}^{N-2}\sum\limits_{r=N-1}^N|\tilde a_{rj}(t)|$ and hence the function $S_{N-1}(t)\dfb (M_{N-1}(t)-M_{N-2}(t))\sum\limits_{j=1}^{N-2}\sum\limits_{r=N-1}^N|\tilde a_{jr}(t)|$ belongs to $L^1$. If $\tilde\eta_{N-2,N-1}(t)>0$, one has $\tilde\eta_{N-2,N-1}(t)=|\tilde a_{N-2,N-1}(t)|(M_{N-1}-M_{N-2}(t))$. Therefore, $\tilde\eta_{N-2,N-1}^+\le S_{N-1}$.
Analogously, if $\tilde\eta_{N-2,N}(t)>0$ then $\tilde\eta_{N-2,N}(t)=|\tilde a_{N-2,N}(t)|(M_{N}(t)-M_{N-1}(t)+M_{N-1}-M_{N-2}(t))$ and hence $\tilde\eta_{N-2,N}^+\le S_{N-1}+S_{N}$. Thus
$\tilde\eta_{N-2,N}^+,\tilde\eta_{N-2,N-1}^+\in L^1[0;\infty]$. Applying  \eqref{eq.deriv_mod1}, we get
\ben
\begin{split}
\dot M_{N-2}(t)=-\tilde\eta_{N-2,N-1}^-(t)-\tilde\eta_{N-2,N}^-(t)+\sum_{j=1}^{N-3}\tilde\eta_{N-2,j}(t)+\\
+\tilde\eta_{N-2,N-1}^+(t)+\tilde\eta_{N-2,N}^+(t).
\end{split}
\een
Since $-\tilde\eta_{N-2,N}^-(t)\le 0,-\tilde\eta_{N-2,N-1}^-(t)\le 0$ and $\tilde\eta_{N-2,j}(t)\le 0$ for $j<N-2$, then either all of these functions are summable or $\int_0^{\infty}\dot M_{N-2}(t)dt=-\infty$. The latter is impossible since $M_{N-2}(t)\ge 0$. Therefore, $\tilde\eta_{N-2,j}\in L^1[0;\infty]\,\forall j$.

Applying the same procedure, one proves that $S_{N-2}(t)\dfb (M_{N-2}(t)-M_{N-3}(t))\sum\limits_{j=1}^{N-2}\sum\limits_{r=N-2}^N|\tilde a_{jr}(t)|$ is summable and  $\tilde\eta_{N-3,j}\in L^1[0;\infty]\,\forall j$, and so on, $\tilde\eta_{ij}\in L^1[0;\infty]$.
\end{IEEEproof}
\begin{corollary}\label{cor.limits}
For any solution of \eqref{eq.proto0} one has $\eta_{ij}:=|a_{ij}|(\theta_{ij}\chi_j-\chi_i)\in L^1[0;\infty]$ and $\dot\chi_i\in L^1[0;\infty]$, so the finite limits
$\chi_i^0=\lim\limits_{t\to+\infty} \chi_i(t)$ and $x_i^0:=\lim\limits_{t\to+\infty} x_i(t)=\pm \chi_i^0$ exist. If $(i,j)\in\E[A(\cdot)]$ then $|x_i^0|=|x_j^0|$;
moreover, $x_i^0=x_j^0$ when $(i,j)\in\E^+[A(\cdot)]$ and $x_i^0=-x_j^0$ when $(i,j)\in\E^-[A(\cdot)]$; as a consequence, $x_i^0=x_j^0=0$ if $(i,j)\in\E^+[A(\cdot)]\cap \E^-[A(\cdot)]$.
\end{corollary}
\begin{IEEEproof}
Since $k^1,\ldots,k^N$ is just a permutation of the set $1:N$, we have $\sum_{i,j}|\eta_{ij}|=\sum_{i,j}|\tilde\eta_{ij}|$, and hence $\eta_{ji}\in L^1[0;\infty]$ for any $i,j$. From \eqref{eq.deriv_mod} one immediately obtains that $\dot\chi_i\in L^1[0;\infty]$ for any $i$, from where the existence of the finite limits $\chi^0_i:=\lim\limits_{t\to+\infty} \chi_i(t)$ is immediate. The limits $x_i^0=\lim\limits_{t\to+\infty} x_i(t)=\pm\chi_i^0$ exist since $x_i$ are continuous. From $|\eta_{ij}(t)|\ge|a_{ij}(t)||\chi_j(t)-\chi_i(t)|$ we know that if $\delta:=|\chi_i^0-\chi_j^0|>0$, for large $t>0$ one has $|\eta_{ij}(t)|\ge|a_{ij}(t)|\delta/2$ and thus
$\int_0^{\infty}|a_{ij}(t)|dt<\infty$ ($a_{ij}$ are locally bounded). Therefore, $\chi_i^0=\chi_j^0$ whenever $i$ and $j$ essentially interact. Suppose that $x_i^0=x_j^0\ne 0$. As $t\to\infty$, one has $\sgn x_i(t)=\sgn x_i^0=\sgn x_j(t)$
and thus if $a_{jk}(t)<0$, we have $\eta_{ij}(t)=|a_{ij}(t)|(-x_i(t)-x_j(t))$. Thus $|\eta_{ij}(t)|\ge |a_{ij}^-(t)||x_i^0|$ for $t>0$ sufficiently large, from where one has that
$\int_{ij}|a_{ij}^-(t)|dt<\infty$. So if the agents essentially compete, the option $x_i^0=x_j^0\ne 0$ is impossible, and thus $x_i^0=-x_j^0$ (with possibility of $x_i^0=x_j^0=0$).
Analogously, one can easily show that $|\eta_{ij}(t)|\ge |a_{ij}^+(t)||x_i^0|$ for $t>0$ sufficiently large if $x_i^0=-x_j^0\ne 0$ which proves the essential cooperation excludes the possibility
of $x_i^0=-x_j^0\ne 0$ and thus $x_i^0=x_j^0$. At last, simultaneous essential cooperation and essential competition imply that $x_i^0=x_j^0=-x_j^0=0$.
\end{IEEEproof}
\begin{corollary}\label{cor.mod_conse}
If the network topology is ESC, the protocol \eqref{eq.proto0} establishes modulus consensus ($\chi_1^0=\ldots=\chi_N^0$), and $\chi_j^0=0$ unless $\G^{\pm}$ is well-defined ($\E^+\cap\E^-=\emptyset$) and SB.
\end{corollary}
\begin{IEEEproof}
The first statement immediately follows from Corollary~\ref{cor.limits} since for any path $i_1,i_2,\ldots,i_r$ in $\G$ one has $\chi_{i_1}^0=\ldots=\chi_{i_r}^0$. Assume that for some initial vector $x(0)$ one has $\chi_i^0\ne 0$. Corollary~\ref{cor.limits} implies that no pair of agents may be both essentially cooperative and essentially competitive, and thus $\E^+\cap\E^-=\emptyset$ so the signed graph $\G^{\pm}$ is well-defined. We have to show $\G^{\pm}$ is structurally balanced, in other words \cite{Altafini:2013,EasleyKleinberg}, has no negative cycles. Indeed, the weight of any arc $(i,j)\in\E$ is $s_{ij}=1$
if $(i,j)\in\E^+$ (hence $x_i^0=x_j^0$) and $s_{ij}=-1$ if $(i,j)\in\E^-$ (and thus $x_i^0=-x_j^0$) so that $x_i^0x_j^0s_{ij}\ge 0$. Given a cycle $i_1,\ldots,i_n=i_1$,
Multiplying the inequalities $x_{i_k}^0x_{i_{k+1}}^0s_{i_ki_{k+1}}\ge 0$, where $i_1,i_2,\ldots,i_n=i_1$ is a cycle and $k\in 1:(n-1)$, one has
$s_{i_1i_2}s_{i_2i_3}\ldots s_{i_{n-1}i_n}(x_1^0\ldots x_n^0)^2\ge 0$ which means that the cycle is positive.
\end{IEEEproof}

\begin{IEEEproof}[Proof of Theorem~\ref{thm.nontrivial}] The necessity in the first statement part follows from Lemma~\ref{lem.eqsc} and Corollary~\ref{cor.mod_conse}: indeed, the ESC condition is necessary for bipartite consensus independent of the cut-balance property. Under the ESC condition, bipartite consensus is possible only when $\G^{\pm}$ exists and is SB.
To prove sufficiency, suppose that $\G^{\pm}$ is well-defined, strongly connected and structurally balanced.
Thanks to Corollary~\ref{cor.mod_conse}, the protocol \eqref{eq.proto0} establishes modulus consensus, and it remains to show it is bipartite consensus. Indeed, consider the protocol \eqref{eq.proto1}, where $\mathfrak A(t)=(\a_{jk}(t))$ and $\a_{jk}(t)=a_{jk}^+(t)$ when $(j,k)\in\E^+$, $\a_{jk}(t)=a_{jk}^-(t)$ when $(j,k)\in\E^-$, and otherwise $\a_{jk}\equiv 0$. The protocol \eqref{eq.proto1} is equivalent to the protocol \eqref{eq.proto0} and hence establishes modulus consensus of the same type as \eqref{eq.proto1} by Lemma~\ref{lem.omega} and Remark~\ref{rem.omega}.
Let $V_1,V_2$ be the hostile camps of the graph $\G^{\pm}$, covering all its nodes. Taking $x_j=+1$ for $j\in V_1$ and $x_j=-1$ for $j\in V_2$, one can show that $x=(x_1,\ldots,x_N)^T$ is an equilibrium point for \eqref{eq.proto1} and hence bipartite consensus is reached. Accordingly to Corollary~\ref{cor.limits}, $x_i^0=x_j^0$ if $i,j\in V_r^1$ or $i,j\in V_r^2$, and $x_i^0=-x_j^0$ whenever $i\in V_r^1,j\in V_r^2$; therefore, opinions polarize unless $E_r^-=\emptyset$. The claim about stability also follows from Corollary~\ref{cor.limits}. 
\end{IEEEproof}

\begin{remark}
It may seem that Theorem~\ref{thm.nontrivial} may be proved by applying the result for unsigned graphs \cite{TsiTsi:13} to the system, obtained from the protocol \eqref{eq.proto1} just constructed
via the gauge transformation (as Lemma~\ref{lem.fix_division} was derived from the relevant result on cooperative agents). Unfortunately, this is not the case. The problem is that the graph $G[\mathfrak A(\cdot)]$ is no longer cut-balanced. This property depends not only on the integrals $\int_0^{\infty}|a_{jk}(s)|ds$, but on the whole function $A(t)$, and is lost after removing inessential interactions.
\end{remark}

\begin{IEEEproof}[Proof of Theorem~\ref{thm.disconn}] The first and the last claims follow from Corollary~\ref{cor.limits}. The second claim is proved by passing to an auxiliary protocol \eqref{eq.proto1}, constructed in the proof of Theorem~\ref{thm.nontrivial}. Let $V_r^1$ and $V_r^2$ be hostile camps in $\G_r^{\pm}$, then taking $x_i=+1$ for $i\in V_r^1$, $x_i=-1$ for $i\in V_r^2$ and $x_i=0$; otherwise, we get an equilibrium point of \eqref{eq.proto1}, and therefore, modulus consensus in the subcommunity $V_r$ is bipartite. Thanks to Corollary~\ref{cor.limits}, $x_i^0=x_j^0$ if $i,j\in V_r^1$ or $i,j\in V_r^2$, and $x_i^0=-x_j^0$ whenever $i\in V_r^1,j\in V_r^2$; therefore, opinions polarize unless $E_r^-=\emptyset$, when bipartite consensus is established.
\end{IEEEproof}

\subsection{Proof of Theorem~\ref{thm.usc} and Corollary~\ref{cor.conse}}\label{subsec.usc}

We start with some useful estimates for the solutions.
\begin{lemma}\label{lem.tech3}
Suppose that $A:[t_0;t_1]\to\R^N$ is a matrix-valued function,
$|A(t)|_{\infty}\le R$ for a.a. $\in [t_0;t_1]$ and $\theta_0\dfb
e^{-R(t_1-t_0)}$. For any solution of \eqref{eq.proto0} one has
\be\label{eq.theta} |x_k(t)|\le
\theta_0|x_k(t_0)|+(1-\theta_0)|x(t_0)|_{\infty}\,\forall k\in
1:N\,\forall t\in [t_0;t_1]. \ee
\end{lemma}
\begin{IEEEproof}
Let $M\dfb |x(t_0)|_{\infty}$. Thanks to Lemma~\ref{lem.bound}, one has $|x_j(t)|\le M$ for any $j$, $t$. Let $s_k(t)\dfb\sum_{j=1}^N|a_{kj}(t)|$ and $S_k(t)\dfb\int_{t_0}^ts_k(\xi)\,d\xi$.
By assumption, $s_k(t)\le R$ and hence $S_k(t)\le R(t_1-t_0)\,\forall t\in [t_0;t_1]$. From \eqref{eq.deriv_mod} one derives that
\be\label{eq.cauchy}
\chi_k(t)\le \left[\chi_k(t_0)+\int\limits_0^te^{S_k(\tau)}\sum_{i=1}^N|a_{ki}(\tau)|\chi_i(\tau)d\tau\right]e^{-S_k(t)}.
\ee
Since $\chi_i(\tau)\le M$, one has $\sum |a_{ki}(\tau)|\chi_i(\tau)\le Ms_k(\tau)$ and $\chi_k(t)\le M+[\chi_k(t_0)-M]e^{-S_k(t)}\le M+[\chi_k(t_0)-M]\theta_0$.
\end{IEEEproof}

\begin{lemma}\label{lem.tech4}
Under assumptions of Lemma~\ref{lem.tech3}, suppose the graph $G[\int_{t_0}^{t_1}\abs A(t)dt]$ is strongly $\ve$-connected and $\theta\dfb\ve\theta_0^2<1$.
Let $V\subsetneq 1:N$ be a non-empty set, $|x(t_0)|_{\infty}=M$ and $\max\limits_{j\not\in V}|x_j(t_0)|=M'$. Then there exists
$k\in V$ such that
\be\label{eq.theta1}
|x_k(t_1)|\le \theta M'+(1-\theta)M.
\ee
\end{lemma}
\begin{IEEEproof}
Since the graph $G[\int_{t_0}^{t_1}\abs A(t)dt]$ is strongly $\ve$-connected, there exist $i\not\in V$ and $k\in V$ such that $\int_{t_0}^{t_1}|a_{ki}(t)|dt\ge\ve$. From \eqref{eq.theta} one has $\chi_j(t)\le M+(M'-M)\theta\,\forall j\not\in V$ and $\chi_j(t)\le M\,\forall j\in V$, which entails that $e^{S_k(\tau)}\sum_{i=1}^N|a_{ki}(\tau)|\chi_i(\tau)\le Me^{S_k(\tau)}s_k(\tau)-(M-M')\theta\ve$ (we used that $e^{S_k(\tau)}\ge 1$). Inequality \eqref{eq.cauchy} yields now that
$$
\chi_k(t_1)\le M+(M'-M)\theta_0\ve e^{-S_k(t_1)}\le M+(M'-M)\theta,
$$
since $e^{S_k(t_1)}\le e^{R(t_1-t_0)}=\theta_0$.
\end{IEEEproof}

\begin{IEEEproof}[Proof of Theorem~\ref{thm.usc}] By assumption, there exist $R,\ve,T>0$ such that $|A(t)|_{\infty}\le R$ and the graph $G_t=G[\int_{t}^{t+T}\abs A(\tau)d\tau]$ is strongly $\ve$-connected for any $t\ge 0$. Let $\theta_0=e^{-RT}$ and $\theta=\ve\theta_0^2$, without loss of generality we may assume that $\theta<1$.

We know from Lemma~\ref{lem.bound} that the maximal modulus $M_N(t)$ always has a limit: $M_N(t)\to M_*$ as $t\to+\infty$ (here we use the notation introduced in Subsection~\ref{subsec.proof.3} so that $M_j(t)$ is the $j$-th modulus in the ascending order, $M_j(t)=|x_{k^j(t)}(t)|$). Our goal is to show that $M_j(t)\to M_*$ via induction by $j=N,N-1,\ldots,1$ . If $M_*=0$ the latter claim is trivial since $0\le M_j(t)\le M_N(t)$; hence we may assume that $M_*>0$.

For $j=N$ our claim holds by definition of $M_*$. Suppose we have proved that $M_N(t),M_{N-1}(t),\ldots,M_{r+1}(t)\to M_*$ as $t\to+\infty$ and have to show that $M_r(t)\to M_*$. Since $M_r(t)\le M_{r+1}(t)$, it suffices to show that $\varliminf\limits_{t\to+\infty} M_r(t)\ge M_*$.
Assume on the contrary that $\varliminf\limits_{t\to+\infty} M_r(t)=m<M_*$. For any $\delta>0$, we have $M_*+\delta>M_j(t)>M_*-\delta$ for large $t>0$ and $j>r$.
On the other hand, there exist a sequence $t_n\to\infty$, along which $M_r(t_n)<m+\delta$. Assume that $\delta>0$ is so small that $(m+\delta)\theta+(M_*+\delta)(1-\theta)<M_*-\delta$ and
$(m+\delta)\theta_0+(M_*+\delta)(1-\theta_0)<M_*-\delta$, that is, $2\delta<(M_*-m)\max(\theta_0,\theta)$.

Accordingly to Lemma~\ref{lem.tech4}, applied for $t_0=t_n$ and $V=\{k^{r+1}(t_n),\ldots,k^N(t_n)\}$, $M=M_*+\delta$ and $M'=m_*+\delta$, there exists $j\in V$ such that $|x_{j}(t_0+T)|<M_*-\delta$. Similarly, for any $j\not\in V$ we have $|x_j(t_n+T)|<M_*-\delta$ thanks to Lemma~\ref{lem.tech3}. Therefore, at time $t_n+T$  there are at least $r+1$ agents, whose opinions have moduli less than $M_*-\delta$ and hence $M_{r+1}(t_n+T)\le M_*-\delta$ for any $n$. We arrived at the contradiction with the induction hypothesis. We thus proved that $M_j(t)\to M_*\,\forall j$ for any solution of \eqref{eq.proto0}, that is, modulus consensus is established.
\end{IEEEproof}

\begin{IEEEproof}[Proof of Corollary~\ref{cor.conse}]
By virtue of Theorem~\ref{thm.usc}, modulus consensus is established. According to Lemma~\ref{lem.struct} it only three types of such a consensus are possible, which are stability, polarization and consensus. The common feature of the first two types is that
for a.a. $x(0)$ there exists $i\in 1:N$ such that $\lim\limits_{t\to+\infty} x_i(t)\le 0$.
It is well known \cite{Moro:04,LinFrancis:07,MatvPro:2013} that the convex hull of the agents' states $\Delta(t)=[\min\limits_i x_i(t),\max\limits_i x_i(t)]$ is non-expanding over time. Hence if $x_i(0)\ge 1\,\forall i$, then $x_i(t)\ge 1$ for any $t\ge 0$, so the first two options are not possible.
\end{IEEEproof}

\subsection{Proof of Lemma~\ref{lem.tech-nonlin} and Theorems~\ref{thm.nonlin2},~\ref{thm.nonlin1}}\label{subsec.nlin}

We start with proof of Lemma~\ref{lem.tech-nonlin}, being a basis for Theorem~\ref{thm.nonlin2}.

\begin{IEEEproof}[Proof of Lemma~\ref{lem.tech-nonlin}]
We consider system \eqref{eq.nonlin2}, and the protocol \eqref{eq.nonlin3} may be studied in the same way. Equation \eqref{eq.tech2} is immediate from the definitions of $\a_{ij}$ and $H_{ij}$.
As follows from Lemma~\ref{lem.bound}, the solutions of \eqref{eq.tech2} remain bounded since $|x(t)|_{\infty}\le|x(0)|_{\infty}$.
Since $H_{ij}(y,z)>0$ are continuous functions and the set $\{(y,z):|y|,|z|\le |x(0)|_{\infty}\}$ is compact, there exist $M>m>0$ such that
$m\le H_{ij}[y,z]\le M$ whenever $|y|,|z|\le |x(0)|_{\infty}$.
By substituting  $y:=x_j(t)\sgn a_{ij}(t)$ and $z:=x_i(t)$, one shows that $m|a_{jk}|\le|\a_{jk}|\le M|a_{jk}|$, from where the claim of Lemma~\ref{lem.tech-nonlin} is obvious.
\end{IEEEproof}

Now we proceed with the proofs of Theorems~\ref{thm.nonlin2} and \ref{thm.nonlin1}.

\begin{IEEEproof}[Proof of Theorem~\ref{thm.nonlin2}]
Since the right-hand sides of \eqref{eq.nonlin2}, \eqref{eq.nonlin3} are smooth in $x$, the solutions exist locally and are unique. According to Lemma~\ref{lem.bound} and \eqref{eq.tech2}, the solutions remain bounded and thus infinitely prolongable. Under the USC assumption, modulus consensus follows from Theorem~\ref{thm.usc} and Lemma~\ref{lem.tech-nonlin}. If the graph $G[A(\cdot)]$
is ESC and cut-balanced, modulus consensus is implied by Theorem~\ref{thm.nontrivial}.
\end{IEEEproof}

\begin{IEEEproof}[Proof of Theorem~\ref{thm.nonlin1}]
Using Lemma~\ref{lem.bound}, one proves that the solution is bounded and hence its derivative also remains bounded due to \eqref{eq.auxbound},
so any solution is infinitely prolongable. Since $x(t)$ is bounded, $A(t)$ is also bounded due to \eqref{eq.auxbound}. The remaining claims follow now from Theorems~\ref{thm.usc},~ \ref{thm.nontrivial}.
\end{IEEEproof}


\section{Conclusions and related works}

In the present paper, we extend a model of opinion dynamics in
social networks with both attractive and repulsive interactions
between the agents, which was proposed in recent papers by C.
Altafini,  who considered the conventional first-order consensus
protocols over signed graphs. Altafini showed, in particular, the
possibility of opinion polarization if the interaction graph is
structurally balanced. In general, the protocol establishes modulus
consensus, where the agents agree in modulus but may differ in signs
(which not excludes convergence of all opinions to zero). In the
present paper, we have examined dynamics of Altafini's protocols
with switching directed topologies and offer sufficient conditions
for reaching modulus consensus that boil down to uniform strong
connectivity of the network. Moreover, under the assumption of
cut-balance, the uniform connectivity may be further relaxed. In
this case, we have obtained necessary and sufficient conditions for
modulus consensus, classified into stability (converges of opinions
to zero) and bipartite consensus (consensus or polarization).
Getting rid of the restriction of static topologies allows to
examine linear and nonlinear dynamics of social networks, where the
agents may change their relations from friendship to hostility and
vice versa. We are currently working with sociologists to test the
theoretical results presented in this paper using data from human
social groups.

\bibliographystyle{IEEEtran}
\bibliography{consensus}

\begin{thebibliography}{10}
\providecommand{\url}[1]{#1}
\csname url@samestyle\endcsname
\providecommand{\newblock}{\relax}
\providecommand{\bibinfo}[2]{#2}
\providecommand{\BIBentrySTDinterwordspacing}{\spaceskip=0pt\relax}
\providecommand{\BIBentryALTinterwordstretchfactor}{4}
\providecommand{\BIBentryALTinterwordspacing}{\spaceskip=\fontdimen2\font plus
\BIBentryALTinterwordstretchfactor\fontdimen3\font minus
  \fontdimen4\font\relax}
\providecommand{\BIBforeignlanguage}[2]{{%
\expandafter\ifx\csname l@#1\endcsname\relax
\typeout{** WARNING: IEEEtran.bst: No hyphenation pattern has been}%
\typeout{** loaded for the language `#1'. Using the pattern for}%
\typeout{** the default language instead.}%
\else
\language=\csname l@#1\endcsname
\fi
#2}}
\providecommand{\BIBdecl}{\relax}
\BIBdecl

\bibitem{RenBeardBook}
W.~Ren and R.~Beard, \emph{Distributed consensus in multi-vehicle cooperative
  control: theory and applications}.\hskip 1em plus 0.5em minus 0.4em\relax
  Springer, 2008.

\bibitem{MesbahiEgerBook}
M.~Mesbahi and M.~Egerstedt, \emph{Graph Theoretic Methods in Multiagent
  Networks}.\hskip 1em plus 0.5em minus 0.4em\relax Princeton and Oxford:
  Princeton University Press, 2010.

\bibitem{RenCaoBook}
W.~Ren and Y.~Cao, \emph{Distributed Coordination of Multi-agent
  Networks}.\hskip 1em plus 0.5em minus 0.4em\relax Springer, 2011.

\bibitem{DeGroot}
M.~DeGroot, ``Reaching a consensus,'' \emph{Journal of the American Statistical
  Association}, vol.~69, pp. 118--121, 1974.

\bibitem{Moro:05}
L.~Moreau, ``Stability of multiagent systems with time-dependent communication
  links,'' \emph{IEEE Trans. Autom. Control}, vol.~50, no.~2, pp. 169--182,
  2005.

\bibitem{LinFrancis:07}
Z.~Lin, B.~Francis, and M.~Maggiore, ``State agreement for continuous-time
  coupled nonlinear systems,'' \emph{SIAM Journ. of Control and Optimization},
  vol.~46, no.~1, pp. 288--307, 2007.

\bibitem{CaoMorse:08}
M.~Cao, A.~Morse, and B.~Anderson, ``Reaching a consensus in a dynamically
  changing environment: convergence rates, measurement delays, and asynchronous
  event,'' \emph{SIAM Journal of Control and Optimization}, vol.~47, no.~2, pp.
  601--623, 2008.

\bibitem{Blondel:05}
V.~Blondel, J.~Hendrickx, A.~Olshevsky, and J.~Tsitsiklis, ``Convergence in
  multiagent coordination, consensus, and flocking,'' in \emph{Proc. IEEE Conf.
  Decision and Control}, 2005, pp. 2996 -- 3000.

\bibitem{MatvPro:2013}
A.~Matveev, I.~Novinitsyn, and A.~Proskurnikov, ``Stability of continuous-time
  consensus algorithms for switching networks with bidirectional interaction,''
  in \emph{Proceedings of European Control Conference ECC-2013}, 2013, pp.
  1872--1877.

\bibitem{TsiTsi:13}
J.~Hendricx and J.~Tsitsiklis, ``Convergence of type-symmetric and cut-balanced
  consensus seeking systems,'' \emph{IEEE Trans. Autom. Control}, vol.~58,
  no.~1, pp. 214--218, 2013.

\bibitem{Muenz:11}
U.~M\"unz, A.~Papachristodoulou, and F.~Allg\"ower, ``Consensus in multi-agent
  systems with coupling delays and switching topology,'' \emph{IEEE Trans.
  Autom. Control}, vol.~56, no.~12, pp. 2976--2982, 2011.

\bibitem{ShiJohansson:13}
G.~Shi and K.~Johansson, ``Robust consensus for continuous-time multi-agent
  dynamics,'' \emph{SIAM J. Control Optim}, vol.~51, no.~5, pp. 3673--3691,
  2013.

\bibitem{ShiJohansson:13-1}
------, ``The role of persistent graphs in the agreement seeking of social
  networks,'' \emph{IEEE J. On Selected Areas In Communications}, vol.~31,
  no.~9, pp. 595--606, 2013.

\bibitem{Sepul:09}
L.~Scardovi and R.~Sepulchre, ``Synchronization in networks of identical linear
  systems,'' \emph{Automatica}, vol.~45, no.~11, pp. 2557--2562, 2009.

\bibitem{HuZheng:2014}
J.~Hu and W.~Zheng, ``Emergent collective behaviors on coopetition networks,''
  \emph{Physica A}, vol. 378, pp. 1787--1796, 2014.

\bibitem{HuZhu:15}
J.~Hu and H.~Zhu, ``Adaptive bipartite consensus on coopetition networks,''
  \emph{Physica D}, vol. 307, pp. 14--21, 2015.

\bibitem{EasleyKleinberg}
D.~Easley and J.~Kleinberg, \emph{Networks, Crowds and Markets. Reasoning about
  a Highly Connected World}.\hskip 1em plus 0.5em minus 0.4em\relax Cambridge:
  Cambridge Univ. Press, 2010.

\bibitem{WassermanFaust}
S.~Wasserman and K.~Faust, \emph{Social Network Analysis: Methods and
  Applications}.\hskip 1em plus 0.5em minus 0.4em\relax Cambridge: Cambridge
  Univ. Press, 1994.

\bibitem{Flache:2011}
A.~Fl\"ache and M.~Macy, ``Small worlds and cultural polarization,''
  \emph{Journal of Math. Sociology}, vol.~35, no. 1--3, pp. 146--176, 2011.

\bibitem{Dillard:2005}
J.~Dillard and L.~Shen, ``On the nature of reactance and its role in persuasive
  health communication,'' \emph{Communication Monographs}, vol.~72, no.~2, pp.
  144--168, 2005.

\bibitem{HovlandBook}
C.~Hovland, I.~Janis, and H.~Kelley, \emph{Communication and persuasion}.\hskip
  1em plus 0.5em minus 0.4em\relax New Haven: Yale Univ. Press, 1953.

\bibitem{Aronson:2010}
E.~Aronson, T.~Wilson, and R.~Akert, \emph{Social psychology}.\hskip 1em plus
  0.5em minus 0.4em\relax Upper Saddle River, NJ: Prentice Hall, 2010.

\bibitem{XiaCao:11}
W.~Xia and M.~Cao, ``Clustering in diffusively coupled networks,''
  \emph{Automatica}, vol.~47, no.~11, pp. 2395--2405, 2011.

\bibitem{Krause:2002}
R.~Hegselmann and U.~Krause, ``Opinion dynamics and bounded confidence models,
  analysis, and simulation,'' \emph{Journal of Artifical Societies and Social
  Simulation (JASSS)}, vol.~5, no.~3, p.~2, 2002.

\bibitem{DeffuantWeisbuch:2000}
G.~Deffuant, D.~Neau, F.~Amblard, and G.~Weisbuch, ``Mixing beliefs among
  interacting agents,'' \emph{Advances in Complex Systems}, vol.~3, pp. 87--98,
  2000.

\bibitem{Dandekar:2013}
P.~Dandekar, A.~Goel, and D.~Lee, ``Biased assimilation, homophily, and the
  dynamics of polarization,'' \emph{PNAS}, vol. 110, no.~15, pp. 5791--5796,
  2013.

\bibitem{Altafini:2012}
C.~Altafini, ``Dynamics of opinion forming in structurally balanced social
  networks,'' \emph{PLoS ONE}, vol.~7, no.~6, p. e38135, 2012.

\bibitem{Altafini:2013}
------, ``Consensus problems on networks with antagonistic interactions,''
  \emph{IEEE Trans. Autom. Control}, vol.~58, no.~4, pp. 935--946, 2013.

\bibitem{MengCaoJohansson:2014}
Z.~Meng, G.~Shi, K.~Johansson, M.~Cao, and Y.~Hong, ``Modulus consensus over
  networks with antagonistic interactions and switching topologies,''
  \emph{http://arxiv.org/abs/1402.2766}, 2014.

\bibitem{Romanczuk:12}
P.~Romanczuk and L.~Schimansky-Geier, ``Swarming and pattern formation due to
  selective attraction and repulsion,'' \emph{Interface Focus}, vol.~2, pp.
  746--756, 2012.

\bibitem{YuChenCaoLuZhang:13}
W.~Yu, G.~Chen, M.~Cao, J.~L\"u, and H.~Zhang, ``Swarming behaviors in
  multi-agent systems with nonlinear dynamics,'' \emph{Chaos}, vol.~23, p.
  043118, 2013.

\bibitem{WangXieCao:13}
C.~Wang, G.~Xie, and M.~Cao, ``Forming circle formations of anonymous mobile
  agents with order preservation,'' \emph{IEEE Trans. Autom. Control}, vol.~58,
  no.~12, pp. 3248--3254, 2013.

\bibitem{Asya:15}
A.~Zakhar'eva, A.~S. Matveev, M.~C. Hoy, and A.~V. Savkin, ``Distributed
  control of multiple non-holonomic robots with sector vision and range-only
  measurements for target capturing with collision avoidance,''
  \emph{Robotica}, vol.~33, pp. 385--412, 2015.

\bibitem{Ovchinnikov:14}
K.~Ovchinnikov, A.~Semakova, and A.~Matveev, ``Decentralized multi-agent
  tracking of unknown environmental level sets by a team of nonholonomic
  robots,'' in \emph{Proc. of 6th Int. Congress on Ultra Modern Communication
  and Control Systems (ICUMT)}, 2014, pp. 352--359.

\bibitem{Valcher:14}
M.~Valcher and P.~Misra, ``On the consensus and bipartite consensus in
  high-order multi-agent dynamical systems with antagonistic interactions,''
  \emph{Systems Control Letters}, vol.~66, pp. 94--103, 2014.

\bibitem{ZhangChen:14}
H.~Zhang and J.~Chen, ``Bipartite consensus of linear multi-agent systems over
  signed digraphs: An output feedback control approach,'' in \emph{Proc. of
  IFAC World Congress}, 2014, pp. 2118 -- 2123.

\bibitem{Hendrickx:14}
J.~Hendrickx, ``A lifting approach to models of opinion dynamics with
  antagonisms,'' in \emph{Proc. of IEEE Conference on Decision and Control
  (CDC)}, 2014, pp. 2118 -- 2123.

\bibitem{AltafiniLini:15}
C.~Altafini and G.~Lini, ``Predictable dynamics of opinion forming for networks
  with antagonistic interactions,'' \emph{IEEE Trans. on Autom. Control},
  vol.~60, no.~2, pp. 342--357, 2015.

\bibitem{XiaCaoJohansson:15}
W.~Xia, M.~Cao, and K.~Johansson, ``Structural balance and opinion separation
  in trust--mistrust social networks,'' \emph{IEEE Trans. on Control of
  Networks}, 2015 (accepted).

\bibitem{Smith:1988}
H.~Smith, ``Systems of ordinary differential equations which generate an order
  preserving flow. {A} survey of results.'' \emph{SIAM Review}, vol.~30, pp. 87
  -- 113, 1988.

\bibitem{LinFrancis:05}
Z.~Lin, B.~Francis, and M.~Maggiore, ``Necessary and sufficient graphical
  conditions for formation control of unicycles,'' \emph{IEEE Trans. Autom.
  Control}, vol.~50, no.~1, pp. 121--127, 2005.

\bibitem{ProCao:2014}
A.~Proskurnikov and M.~Cao, ``Opinion dynamics using {A}ltafini's model with a
  time-varying directed graph,'' in \emph{Proceedings of IEEE ISIC 2014 (Part
  of IEEE MSC 2014)}, Antibes, 2014, pp. 849--854.

\bibitem{ProMatvCao:2014}
A.~Proskurnikov, A.~Matveev, and M.~Cao, ``Consensus and polarization in
  {A}ltafini's model with bidirectional time-varying network topologies,'' in
  \emph{Proceedings of IEEE CDC 2014}, Los Angeles, 2014, pp. 2112--2117.

\bibitem{Murray:04}
R.~Olfati-Saber and R.~Murray, ``Consensus problems in networks of agents with
  switching topology and time-delays,'' \emph{IEEE Trans. Autom. Control},
  vol.~49, no.~9, pp. 1520--1533, 2004.

\bibitem{AgaevChe:2014}
P.~Chebotarev and R.~Agaev, ``The forest consensus theorem,'' \emph{IEEE Trans.
  Autom. Control}, vol.~59, no.~9, pp. 2475--2479, 2014.

\bibitem{Moro:04}
L.~Moreau, ``Stability of continuous-time distributed consensus algorithms,''
  in \emph{Proc. IEEE Conf. Decision and Control (CDC 2004)}, 2004, pp. 3998 --
  4003.

\bibitem{Marvel:2011}
S.~Marvel, J.~Kleinberg, R.~Kleinberg, and S.~Strogatz, ``Continuous-time model
  of structural balance,'' \emph{PNAS}, vol. 108, no.~5, pp. 1771--1776, 2011.

\bibitem{Antal:2005}
T.~Antal, P.~Krapivsky, and S.~Redner, ``Dynamics of social balance on
  networks,'' \emph{Phys. Rev. E}, vol.~72, p. 036121, 2005.

\bibitem{AgaevChe:2005}
R.~Agaev and P.~Chebotarev, ``On the spectra of nonsymmetric laplacian
  matrices,'' \emph{Linear Algebra Appl.}, vol. 399, pp. 157--168, 2005.

\bibitem{Clarke:1975}
F.~Clarke, ``Generalized gradients and applications,'' \emph{Trans. of Amer.
  Math. Soc.}, vol. 205, pp. 247 -- 262, 1975.

\bibitem{McShane:1967}
E.~McShane and R.~Warfield, ``On {F}ilippov's implicit functions lemma,''
  \emph{Proc. Amer. Math. Soc.}, vol.~18, pp. 41--47, 1967.

\end{thebibliography}

\begin{IEEEbiography}[{\includegraphics[width=1in,height=1.25in,clip,keepaspectratio]{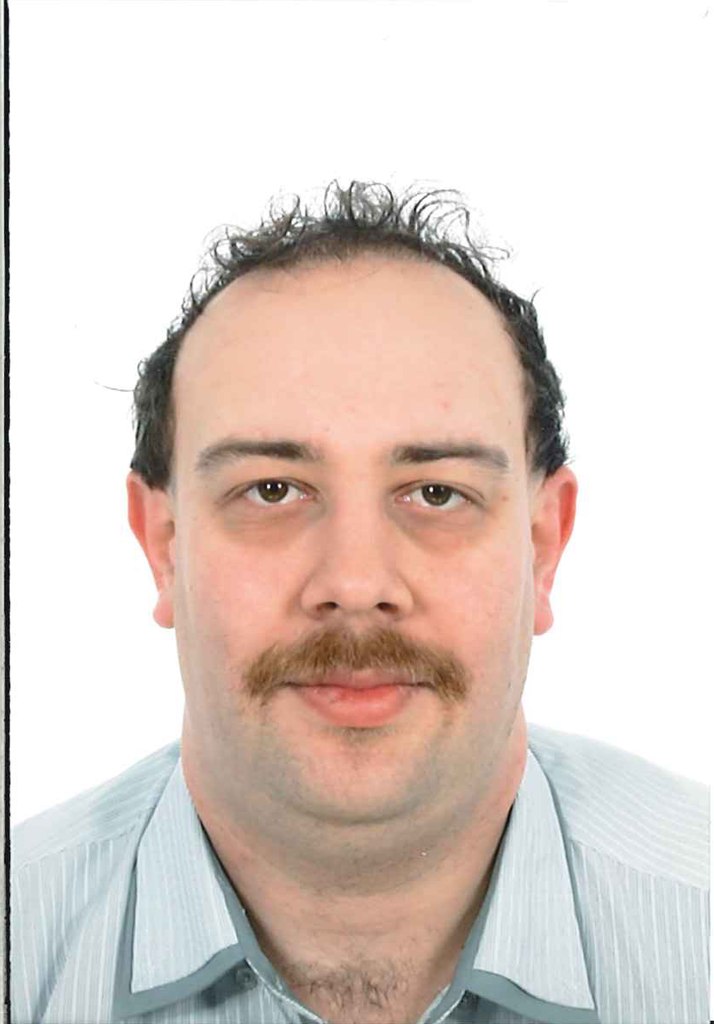}}]{Anton V. Proskurnikov}
was born in Leningrad (now St. Petersburg) in 1982. He received his M.Sc. and Ph.D. degrees in 2003 and 
2005 respectively, both from St.-Petersburg State University and supervised by Prof. V.A. Yakubovich. 
From 2003 till 2010 he was an Assistant Professor of the Department
of Mathematics and Mechanics, St.Petersburg State University.
Now A. Proskurnikov is a postdoctoral researcher at the ENTEG institute the University of Groningen, The Netherlands. 
He also occupies part-time researcher positions at St. Petersburg State University and
Institute for Problems of Mechanical Engineering of Russian Academy of Sciences, St.Petersburg, Russia. 
His research interests include dynamics and control of complex networks, multi-agent and decentralized control nonlinear control,
robust control, optimal control and optimization. In 2009 Anton Proskurnikov was awarded with a medal of Russian Academy of Sciences for prominent 
young researchers for a series of works on optimal tracking and disturbance rejection.
 \end{IEEEbiography}

\begin{IEEEbiography}[{\includegraphics[width=1in,height=1.25in,clip,keepaspectratio]{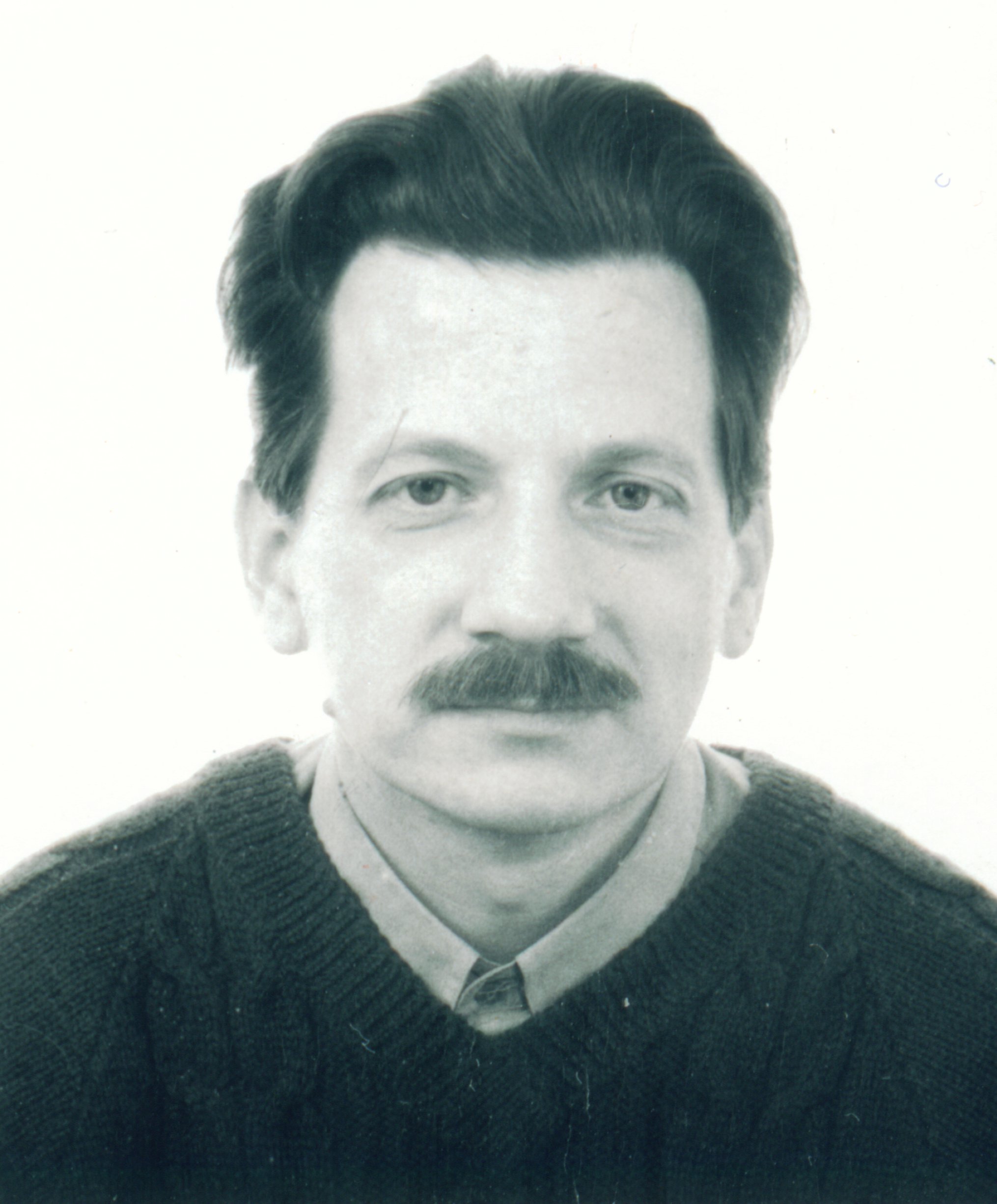}}]{Alexey S. Matveev}
was born in Leningrad, USSR, in 1954. He received the
M.S. and Ph.D. degrees in 1976 and 1980, respectively, both from the
Leningrad University. Currently, he is a professor of the Department
of Mathematics and Mechanics, St. Petersburg State University. His
research interests include estimation and control over communication networks, hybrid dynamical systems, 
and navigation and control of mobile robots.
 \end{IEEEbiography}

\begin{IEEEbiography}[{\includegraphics[width=1in,height=1.25in,clip,keepaspectratio]{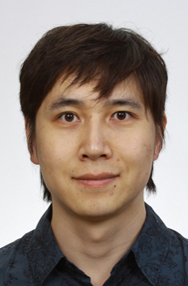}}]{Ming Cao}
 is currently a tenured associate professor responsible for the research direction of network analysis and control 
with the Faculty of Mathematics and Natural Sciences at the University of Groningen, the Netherlands, where he started as a 
tenure-track assistant professor in 2008. He received the Bachelor degree in 1999 and the Master degree in 2002 from Tsinghua 
University, Beijing, China, and the PhD degree in 2007 from Yale University, New Haven, CT, USA, all in electrical engineering. 
From September 2007 to August 2008, he was a postdoctoral research associate with the Department of Mechanical and Aerospace 
Engineering at Princeton University, Princeton, NJ, USA. He worked as a research intern during the summer of 2006 with the 
Mathematical Sciences Department at the IBM T. J. Watson Research Center, NY, USA. His main research interest is in autonomous 
agents and multi-agent systems, mobile sensor networks and complex networks.  He is an associate editor for Systems 
and Control Letters, and for the Conference Editorial Board of the IEEE Control Systems Society. He is also a member of the 
IFAC Technical Committee on Networked Systems.
\end{IEEEbiography}
\end{document}